\documentclass[iop]{emulateapj}
\usepackage{amsmath}

\newcommand{\kms}{\rm km~s^{-1}}
\newcommand{\kmsmpc}{\rm km~s^{-1}~Mpc^{-1}}
\newcommand{\dn}{D_{n}4000}

\usepackage{color}
\usepackage{multirow}

\begin{document}

\title{The Velocity Dispersion Function for Quiescent Galaxies in Nine Strong-Lensing Clusters}

\author{Jubee Sohn$^{1}$,
        Daniel G. Fabricant$^{1}$, Margaret J. Geller$^{1}$, 
        Ho Seong Hwang$^{2}$, Antonaldo Diaferio$^{3,4}$}

\affil{$^{1}$ Smithsonian Astrophysical Observatory, 60 Garden Street, Cambridge, MA 02138, USA}
\affil{$^{2}$ Korea Astronomy and Space Science Institute, 776 Daedeokdae-ro, Yuseong-gu, Daejeon 34055, Korea}
\affil{$^{3}$ Universit\`a di Torino, Dipartimento di Fisica, Torino, Italy}
\affil{$^{4}$ Istituto Nazionale di Fisica Nucleare (INFN), Sezione di Torino, Torino, Italy}

\email{jubee.sohn@cfa.harvard.edu}

%=============================================================
\begin{abstract}
We measure the central stellar velocity dispersion function for quiescent galaxies in a set of nine northern clusters in the redshift range $0.18 < z < 0.29$ and with strong lensing arcs in Hubble Space Telescope images. The velocity dispersion function links galaxies directly to their dark matter halos. From dense SDSS and MMT/Hectospec spectroscopy we identify $231 - 479$ spectroscopic members in each cluster. We derive physical properties of cluster members including redshift, $\dn$, and central stellar velocity dispersion and we include a table of these measurements for 3419 cluster members. We construct the velocity dispersion functions for quiescent galaxies with $\dn > 1.5$ and within $R_{200}$. The cluster velocity dispersion functions all show excesses at $\sigma \gtrsim 250~\kms$ compared to the field velocity dispersion function. The velocity dispersion function slope at large velocity dispersion ($\sigma > 160~\kms$) is steeper for more massive clusters, consistent with the trend observed for cluster luminosity functions. The spatial distribution of galaxies with large velocity dispersion at radii larger than $R_{200}$ further underscores the probable major role of dry mergers in the growth of massive cluster galaxies during cluster assembly. 
\end{abstract}

%=============================================================
\section{INTRODUCTION}

In the standard $\Lambda$CDM paradigm, dark matter (DM) halos grow hierarchically through accretion of neighboring substructures. Within DM halos, galaxies form and evolve through a complicated interplay between the dark and baryonic matter. Hierarchical structure formation naturally leads to environmental effects because the host DM halo mass plays a critical role in galaxy evolution. 

For decades there have been many explorations of environmental effects on galaxy properties including morphology, color, mass, star formation rate, and the ages of their stellar populations (e.g., \citealp{1980ApJ...236..351D, 1999ApJ...527...54B, 2006PASP..118..517B, 2007ApJ...658..898P, 2009ARA&A..47..159B, 2010ApJ...721..193P, 2016ApJ...816L..25P}). At a fixed stellar mass, clusters include quiescent galaxies that are older than their counterparts in lower density environments (e.g., \citealp{2004MNRAS.353..713K, 2009ApJ...699.1595P, 2009ARA&A..47..159B}). 

The luminosity function of galaxies has long been used to link galaxies to their DM subhalos (e.g., \citealp{1999ApJ...516..530K, 2004MNRAS.353..189V, 2008ApJ...676..248Y}). The luminosity function can trace environmentally induced changes in the DM subhalo mass distribution \citep{2004MNRAS.353..189V}. The analogous stellar mass function is another observational tracer of the underlying halo mass distribution (e.g., \citealp{2010ApJ...710..903M, 2018ARA&A..56..435W, 2019MNRAS.488.3143B}). Many studies show that the shapes of luminosity and stellar mass functions vary with environment (e.g., \citealp{2012ApJ...752...64H, 2013MNRAS.432.3141C, 2013A&A...550A..58V, 2016A&A...586A..23D, 2018ApJ...854...30P}). For example, based on the luminosity function for star-forming and quiescent populations at different redshifts, \citet{2010ApJ...721..193P} propose a galaxy evolution model based on a combination of mass quenching and environmental quenching.

For the quiescent galaxy population, the stellar velocity dispersion function complements the luminosity and stellar mass functions (e.g., \citealp{2003ApJ...594..225S, 2010MNRAS.402.2031C}). The stellar velocity dispersion of a quiescent galaxy is a proxy for the dark matter halo dispersion \citep{2012ApJ...751L..44W, 2015ApJ...800..124B, 2016ApJ...832..203Z, 2018ApJ...859...96Z}. The central stellar velocity dispersion is insensitive to the complex physics of the evolving stellar population and to minor merging. Measuring the stellar velocity dispersions is straightforward compared to untangling the systematic errors in dense cluster photometry that affect both the luminosity and stellar mass measurements.

Determination of stellar velocity dispersion functions requires large, dense spectroscopic surveys. Large spectroscopic surveys (e.g., Sloan Digital Sky Survey (SDSS) and Baryon Oscillation Spectroscopic Survey (BOSS)) provide robust measurements of the general field velocity dispersion function for quiescent galaxies \citep{2003ApJ...594..225S, 2005ApJ...622...81M, 2007ApJ...658..884C, 2010MNRAS.402.2031C, 2016MNRAS.461.1131M, 2017ApJ...845...73S, 2019arXiv190400486H}. \citet{2017ApJS..229...20S} measure the velocity dispersion function for quiescent galaxies in the local massive clusters, Coma and A2029. Based on homogeneous quiescent galaxy selection and velocity dispersion measurements, \citet{2017ApJ...845...73S} show that the cluster velocity dispersion function significantly exceeds the field velocity dispersion function for $\sigma > 250~\kms$. This difference between the cluster and field velocity dispersion functions underscores the dependence of galaxy formation and evolution on local environment. 

Here we measure the velocity dispersion function for nine massive clusters at $0.18 < z < 0.29$. This set of northern clusters have strong lensing arcs identified in Hubble Space Telescope images taken before 2015. In each cluster, we use the velocity dispersion of $120 - 267$ spectroscopically identified quiescent members within $R_{200}$ to construct the cluster velocity dispersion function. For the first time, we investigate the shape of the cluster velocity dispersion function as a function of cluster mass. We discuss the results in the context of the current observational and theoretical pictures of galaxy evolution in clusters.

We describe the cluster sample (including a table of the relevant data) along with the photometric and spectroscopic measurements of cluster galaxies in Section \ref{data}. In Section \ref{vdf}, we describe the details of constructing the cluster velocity dispersion functions. We compare the cluster velocity dispersion functions with previously publishes velocity dispersion functions for two local massive clusters, Coma and A2029 \citep{2017ApJS..229...20S} and for quiescent galaxies in the field \citep{2017ApJ...845...73S}. In Section \ref{discussion}, we discuss the interpretation of the cluster velocity dispersion functions based on the previous theoretical and observational work. We conclude in Section \ref{conclusion}. Throughout the paper, we adopt a $\Lambda$CDM cosmology with $H_{0} = 70~\kmsmpc$, $\Omega_{m} = 0.3$, and $\Omega_{\Lambda} = 0.7$. 

%=============================================================
\section{DATA} \label{data}

We first describe the spectroscopic surveys. Then we outline the identification of the quiescent cluster members we use to measure the velocity dispersion functions. We describe the target clusters in Section \ref{target}. Section \ref{phot} and Section \ref{spec} describe the photometric and spectroscopic data and the spectroscopic properties of the cluster members.
 
\subsection{Target Clusters}\label{target}

We examine the galaxy velocity dispersion functions of nine clusters with $0.18 < z < 0.28$. As of 2015, this sample included all northern strong lensing clusters in this redshift range with Hubble Space Telescope (HST) imaging. Our dense spectroscopic surveys of these clusters enabled comparison of strong lensing and spectroscopic measures of the masses of cluster galaxies \citep{2015MNRAS.447.1224M, 2017MNRAS.465.4589M}.
 
Table \ref{list} lists key properties of the target clusters including ID, R.A., Decl., redshift, the number of spectroscopically identified members, the number of members with a central stellar velocity dispersion measurement, $R_{200}$ and $M_{200}$. Here, $R_{200}$ is the radius that contains a mean density 200 times larger than the critical density of the universe, and $M_{200}$ is the mass enclosed within $R_{200}$. We derive the number of spectroscopic members, $R_{200}$ and $M_{200}$ based on the caustic technique (Section \ref{member}, \citealp{1997ApJ...481..633D, 1999MNRAS.309..610D, 2011MNRAS.412..800S, 2013ApJ...768..116S}). 

%=================================
%Table \ref{list}
%=================================
\begin{deluxetable*}{cccccccccc}
\tablenum{1}
\tablecaption{The cluster sample}
\tablecolumns{10}
\tabletypesize{\scriptsize}
\tablewidth{0pt}
\tablehead{
\multirow{2}{*}{ID}  & \colhead{R.A.}  & \colhead{Decl.}  & \multirow{2}{*}{z} &  
\multirow{2}{*}{$N_{mem}$\tablenotemark{$^a$}} &
\multirow{2}{*}{$N_{200}$\tablenotemark{$^b$}} &
\multirow{2}{*}{$N_{200, Q}$\tablenotemark{$^c$}} &
\multirow{2}{*}{$N_{200, Q, \sigma}$\tablenotemark{$^d$}} & 
\colhead{$R_{200}$} & \colhead{$M_{200}$} \\
 					 & \colhead{(deg)} & \colhead{(deg)}  &                    &
 					                           & 
 					                           &
 					                            		  &
\colhead{(Mpc)}	    & \colhead{(10$^{14}$ $M_{\odot}$)} }
\startdata
A2390   & 328.398383 &  17.697349 & 0.22829 &  479 &  227 &  169 &  159 & $2.34 \pm 0.05$ & $18.42 \pm 1.13$ \\
A1703   & 198.766796 &  51.824892 & 0.27762 &  466 &  187 &  160 &  155 & $2.06 \pm 0.10$ & $13.22 \pm 2.04$ \\
A1689   & 197.872924 &  -1.338086 & 0.18424 &  416 &  267 &  218 &  207 & $2.08 \pm 0.21$ & $12.31 \pm 4.17$ \\
A1835   & 210.265371 &   2.879629 & 0.25063 &  364 &  178 &  136 &  132 & $2.02 \pm 0.05$ & $12.02 \pm 0.93$ \\
A773    & 139.474430 &  51.736714 & 0.21728 &  441 &  212 &  164 &  156 & $1.99 \pm 0.01$ & $11.15 \pm 0.16$ \\
A68     &   9.264441 &   9.178039 & 0.25045 &  254 &  137 &  113 &  110 & $1.89 \pm 0.01$ & $ 9.89 \pm 0.18$ \\
A611    & 120.237028 &  36.051250 & 0.28730 &  400 &  178 &  141 &  135 & $1.84 \pm 0.13$ & $ 9.51 \pm 2.12$ \\
RXJ2129 & 322.409370 &   0.084740 & 0.23387 &  368 &  120 &  105 &   99 & $1.77 \pm 0.17$ & $ 7.95 \pm 2.53$ \\
A383    &  42.018376 &  -3.525639 & 0.18849 &  231 &  143 &   97 &   94 & $1.65 \pm 0.03$ & $ 6.15 \pm 0.39$
\enddata
\tablenotetext{a}{The number of spectroscopic members.}
\tablenotetext{b}{The number of spectroscopic members within $R_{200}$.}
\tablenotetext{c}{The number of quiescent members ($\dn > 1.5$) within $R_{200}$.}
\tablenotetext{c}{The number of quiescent members ($\dn > 1.5$) with $\sigma$ measurements within $R_{200}$.}
\label{list}
\end{deluxetable*}
%=================================

\subsection{Photometric Data}\label{phot} 

We selected the photometric sample from the SDSS Data Release (DR) 14 \citep{2018ApJS..235...42A}. SDSS provides a uniform photometric galaxy sample across the $30 \arcmin$ radius Hectospec field of view. The source catalog includes all objects within $40 \arcmin$ of the cluster center. We cross-matched the objects in the SDSS catalog with the DESI (Dark Energy Survey Instrument) Legacy Survey catalog DR8 \citep{2019AJ....157..168D}. For $r < 22$ and $ R < R_{200}$, there are no Legacy Survey matches for $5\%$ of the objects in the SDSS catalogs. The unmatched objects are blends of close stars, high proper motion stars, and artifacts including diffraction spikes. We drop these objects from further consideration. We then define galaxies as objects with either an SDSS $probPSF = 0$ ($probPSF:$ probability that the object is a star) or a Legacy Survey classification other than PSF. We use the Legacy Survey DR8 $r$ model magnitudes and extinctions in the following analysis.

\subsection{Spectroscopic Data}\label{spec}

%=====================================
% Figure \ref{zcomp_mag}
%=====================================
\begin{figure*}
\centering
\includegraphics[scale=0.47]{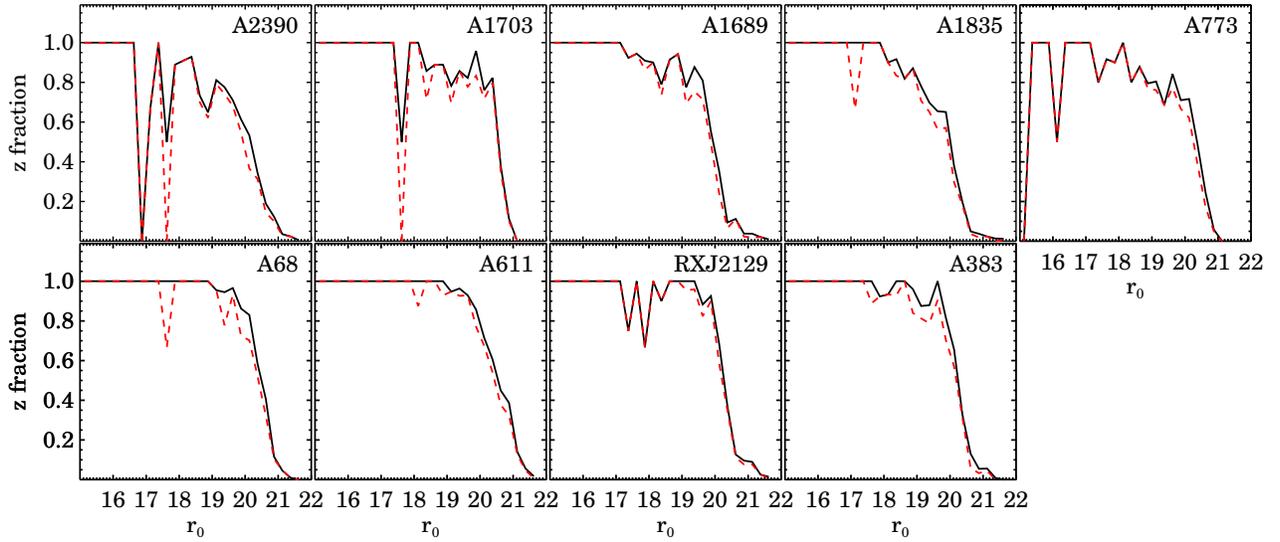}
\caption{Spectroscopic survey completeness as a function of r-band magnitude within $R_{cl} < R_{200}$. Solid lines show the redshift completeness and dashed lines show the velocity dispersion completeness regardless of $\dn$. } 
\label{zcomp_mag}
\end{figure*}
%=====================================

%=====================================
% Figure \ref{zcomp_map}
%=====================================
\begin{figure*}
\centering
\includegraphics[scale=0.47]{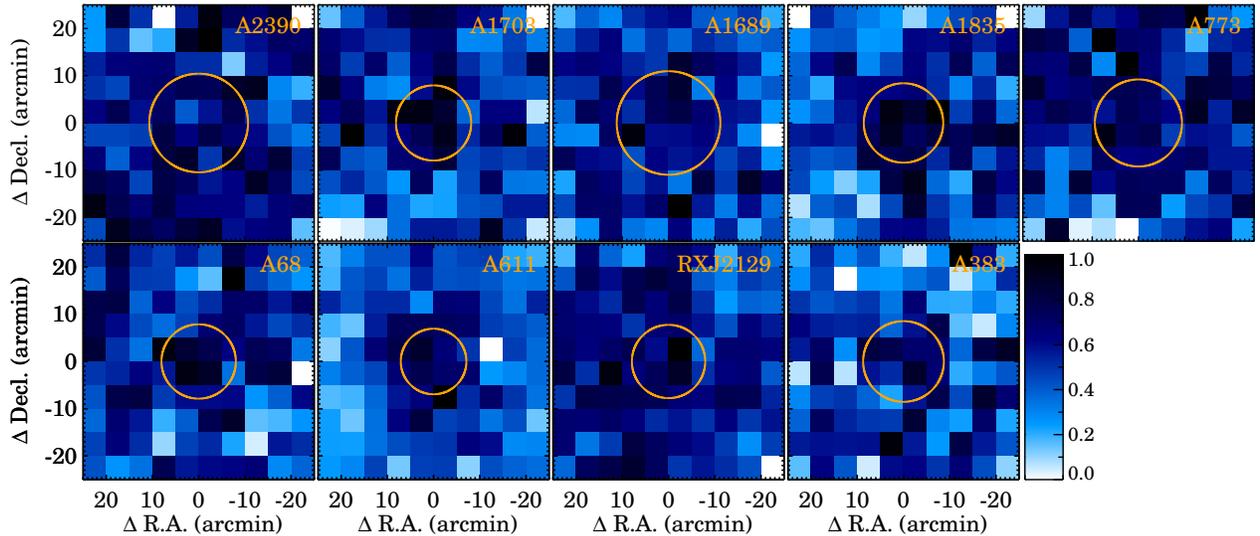}
\caption{Spectroscopic survey completeness map for galaxies brighter than $r = 20$ for each cluster. Darker color indicate higher survey completeness. Solid circles show $R_{200}$ for each cluster. } 
\label{zcomp_map}
\end{figure*}
%=====================================

Spectroscopically identified members with a central stellar velocity dispersion measurement are the first step in the construction of the velocity dispersion function. We compile previous spectroscopic data obtained with three instruments: SDSS \citep{2006AJ....131.2332G}, BOSS \citep{2013AJ....145...10D}, and Hectospec \citep{2005PASP..117.1411F}. We obtain the SDSS and BOSS data from SDSS DR14. 

Hectospec is a 300 fiber-fed spectrograph at the MMT covering a circular field with radius ($R = 30\arcmin$). We compile Hectospec data from two previous surveys: Hectospec Cluster Surveys (HeCS, \citealp{2013ApJ...767...15R, 2016ApJ...819...63R, 2018ApJ...862..172R, 2014ApJ...783...52G, 2014ApJ...797..106H}) and the Arizona Cluster Redshift Survey\footnote{http://herschel.as.arizona.edu/acres/data/acres data.php} (ACReS, \citealp{2015ApJ...806..101H, 2013ApJ...775..126H}). ACReS includes 7 of the 9 clusters in this study: A68, A363, A611, A1689, A1835, A2390, and RXJ 2129. A1703 was not included in either HeCS or ACReS. 

We carried out additional Hectospec observations for all of the target clusters between 2011 and 2017, with the objective of producing a magnitude-limited survey to $r = 20$ within the Hectospec field of view (FoV). We used the 270 line mm$^{-1}$ grating following HeCS and ACReS. Each Hectospec field was observed with three sequential exposures of 1200 s (similar to the HeCS and ACReS). The resulting Hectospec spectra cover $3750 - 9100~{\rm \AA}$ at $R \sim 1000$; SDSS spectra cover $3800 - 9200~{\rm \AA}$ at $R \sim 2000$ and BOSS spectra cover $3560 - 10400~{\rm \AA}$ also at $R \sim 2000$. 

We reduce the Hectospec spectra including HeCS and ACRES spectra using the HSRED v2.0 package\footnote{http://oirsa.cfa.harvard.edu/archive/search/}. We measure redshifts with the cross-correlation package RVSAO \citep{1998ASPC..145...93M}. We select reliable redshifts with $R_{XC} > 5$ where $R_{XC}$ is the cross-correlation score from RVSAO, slightly more conservative than the selection from \citet{2016ApJ...819...63R}. For ACReS objects, we take the redshift from the ACReS catalog and confirm their results by visual inspection. The typical uncertainty of the Hectospec redshifts is $30~\kms$. Table \ref{list} summarizes the number of redshifts within $R < R_{200}$. We also derive $\dn$ \citep{2008PASP..120.1222F} and the central stellar velocity dispersion \citep{2013PASP..125.1362F} from the MMT Hectospec spectra.
 
Figure \ref{zcomp_mag} shows the spectroscopic completeness as a function of $r-$band magnitude for each cluster. We compute the spectroscopic completeness within $R_{200}$, where the mass density is 200 times the critical density at the cluster redshift (see Section \ref{member}). The spectroscopic survey is typically $\gtrsim 65\%$ complete at the survey limit $r = 20$. Figure \ref{zcomp_map} displays a 2D map of the spectroscopic completeness for each cluster to $r = 20$. Darker colors indicate higher spectroscopic completeness.

\subsubsection{Member Identification}\label{member}

We use the caustic technique \citep{1997ApJ...481..633D, 1999MNRAS.309..610D, 2013ApJ...768..116S} to identify spectroscopic members. This technique was originally developed to measure the dynamical mass of galaxy clusters \citep{1997ApJ...481..633D}, but it is also a valuable technique for determining cluster membership. \citet{2013ApJ...768..116S} demonstrate that the caustic technique identifies 95\% of members within $3R_{200}$ of simulated clusters with $\sim 180$ members in fields containing $\sim 1000$ galaxies. Several previous observational studies of galaxy clusters use the caustic technique for membership determination (e.g., \citealp{2006AJ....132.1275R, 2013ApJ...767...15R, 2014ApJ...797..106H, 2016ApJ...819...63R, 2018ApJ...862..172R, 2017ApJS..229...20S, 2019ApJ...872..192S, 2020ApJ...891..129S}). 
 
%=====================================
% Figure \ref{rv}
%=====================================
\begin{figure*}
\centering
\includegraphics[scale=0.47]{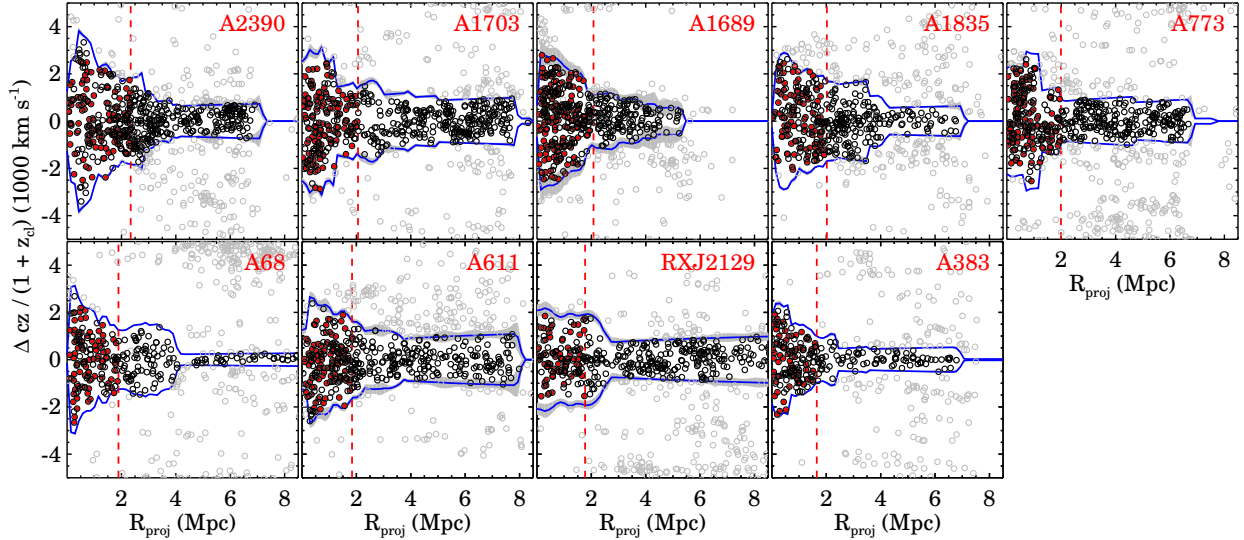}
\caption{
Rest-frame clustercentric radial velocity as a function of projected clustercentric distance. Black open and red filled circles indicate cluster members and the quiescent cluster members with a velocity dispersion measurement, respectively. Gray circles indicate spectroscopically identified non-members. Blue solid lines show the caustics for each target cluster and the shaded regions indicate the uncertainty in the caustics. The vertical dashed lines mark $R_{200}$. } 
\label{rv}
\end{figure*}
%===================================== 
 
Figure \ref{rv} shows the relative galaxy velocity difference in the rest frame with respect to the mean cluster redshift for each cluster as a function of projected clustercentric distance (the R-v diagram). Blue curves indicate the caustic pattern. The galaxies within these boundaries are identified as cluster members (black open circles); red filled circles indicate quiescent cluster members with velocity dispersion measurements. For the set of clusters, there are 120 - 267 spectroscopic members within $R_{200}$ and the mean number of cluster members is 183. 

The caustic technique yields the mass profile as a function of clustercentric distance. Based on this mass profile, we compute $M_{200}$ and $R_{200}$ (see Table \ref{list}). $M_{200}$ ranges from $6.1 \times 10^{14} M_{\odot}$ to $1.8 \times 10^{15} M_{\odot}$ with a mean of $1.1 \times 10^{15} M_{\odot}$; $R_{200}$ ranges from 1.65 Mpc to 2.34 Mpc with a mean of 1.96 Mpc.

We include a catalog of spectroscopically identified members of the target clusters. Table \ref{cat} lists Cluster ID, SDSS object ID, R.A., Decl., redshift, and the $r-$band apparent and absolute magnitudes of the cluster members (not limited by $R_{cl} < R_{200}$, where $R_{cl}$ is the clustercentric distance). We also list spectroscopic quantities including $\dn$ and $\sigma$ (see below) used for deriving the velocity dispersion functions. 

%=================================
%Table \ref{list}
%=================================
\begin{deluxetable*}{lcccccccc}
\tablenum{2}
\tablecaption{Spectroscopic Members \label{cat}}
\tablecolumns{9}
\tabletypesize{\scriptsize}
\tablewidth{0pt}
\tablehead{\colhead{Cluster} & \colhead{SDSS ObjID} & \colhead{R.A.} & \colhead{Decl.} & \colhead{z} & \colhead{r} & \colhead{$M_{r}$} & \colhead{$\dn$} & \colhead{$\sigma$}\\
 & & (deg) & (deg) & & (mag) & (mag) & & ($\kms$)} 
\startdata
A2390 & 1237680297267954250 & 328.311665 & 17.634877 & $0.2262 \pm 0.0001$ &  18.57 & -21.65 & $1.23 \pm 0.03$ & $ 98 \pm  12$ \\
A2390 & 1237680297267954147 & 328.249936 & 17.672913 & $0.2290 \pm 0.0001$ &  18.86 & -21.60 & $1.67 \pm 0.07$ & $134 \pm  23$ \\
A2390 & 1237680297267954148 & 328.253163 & 17.675519 & $0.2249 \pm 0.0001$ &  19.22 & -21.13 & $1.44 \pm 0.07$ & \nodata \\
A2390 & 1237680297268020299 & 328.390579 & 17.536742 & $0.2263 \pm 0.0001$ &  19.53 & -20.97 & $1.89 \pm 0.08$ & $114 \pm  21$ \\
A2390 & 1237680297268020675 & 328.482933 & 17.550532 & $0.2323 \pm 0.0001$ &  19.95 & -20.55 & $2.03 \pm 0.19$ & $ 59 \pm  34$ \\
A2390 & 1237680297268020338 & 328.389801 & 17.582862 & $0.2341 \pm 0.0001$ &  19.90 & -20.60 & $1.66 \pm 0.07$ & $164 \pm  18$ \\
A2390 & 1237680297268020635 & 328.455222 & 17.625013 & $0.2255 \pm 0.0002$ &  19.57 & -20.95 & $1.61 \pm 0.07$ & $132 \pm  28$ \\
A2390 & 1237680297268020399 & 328.387204 & 17.662835 & $0.2168 \pm 0.0002$ &  20.52 & -19.94 & $1.43 \pm 0.16$ & $ 93 \pm  59$ \\
A2390 & 1237680297268020236 & 328.340551 & 17.679817 & $0.2258 \pm 0.0001$ &  19.67 & -20.66 & $1.42 \pm 0.09$ & $178 \pm  48$ \\
A2390 & 1237680297268019264 & 328.357527 & 17.701409 & $0.2387 \pm 0.0002$ &  20.28 & -20.24 & $1.73 \pm 0.12$ & $149 \pm  36$
\enddata
\tablenotetext{}{The entire table is available in machine-readable form in the online journal. Here, a portion is shown for guidance regarding its format.}
\end{deluxetable*}
%=================================

\subsubsection{$D_{n}4000$}

$\dn$ is a useful spectral index that probes the stellar population age (e.g., \citealp{2003MNRAS.341...33K}). Following \cite{1999ApJ...527...54B} we define $\dn$ as the flux ratio between $3850 - 3950 {\rm \AA}$ and $4000 - 4100 {\rm \AA}$. We measure $\dn$ from spectra obtained with the SDSS, BOSS, and Hectospec spectrographs. The $\dn$ values from these spectrographs are consistent within $\sim 5\%$ \citep{2013PASP..125.1362F}. 

A $\dn$ of 1.5 corresponds to a stellar population 1 Gyr old, and increases monotonically as the stellar population ages \citep{2003MNRAS.341...33K}. Thus, $\dn$ is widely used to identify quiescent galaxies. Following previous studies exploring the velocity dispersion of quiescent galaxies \citep{2017ApJS..229...20S, 2017ApJ...845...73S, 2016ApJ...832..203Z}, we select quiescent galaxies with $\dn > 1.5$.

\subsubsection{Central Stellar Velocity Dispersion}\label{sigma}

We obtain velocity dispersion measurements for SDSS and BOSS galaxies from the Portsmouth reduction \citep{2013MNRAS.431.1383T}. These measurements are consistent with the Hectospec measurements \citep{2013PASP..125.1362F}. The Portsmouth reduction derives velocity dispersions using Penalized Pixel-Fitting (pPXF, \citealp{2004PASP..116..138C}). The best-fit velocity dispersions are derived by comparing SDSS spectra with stellar population templates \citep{2011MNRAS.418.2785M} generated from the MILES stellar library \citep{2006A&A...457..809S}. We obtained the Portsmouth velocity dispersion for less than 12 spectroscopic members in each cluster. The typical uncertainty of the Portsmouth velocity dispersion for cluster members is $12~\kms$.

We measure central stellar velocity dispersions from Hectospec spectra using the pPXF-based UlySS code (University of Lyon Spectroscopic analysis Software, \citealp{2009A&A...501.1269K}). We construct stellar population templates based on the MILES stellar library using the PEGASE-HR code \citep{2004A&A...425..881L}. ULySS convolves these templates to the Hectospec resolution at varying velocity dispersions, stellar population ages, and metallicities, and then finds the best-fit age, metallicity, and velocity dispersion through chi-square minimization of the template fits. To minimize the uncertainty in the velocity dispersion \citep{2013PASP..125.1362F}, we derive the fit within the rest-frame spectral range $4100 - 5500~ {\rm \AA}$. The typical uncertainty of the Hectospec velocity dispersion for cluster members is $24~\kms$. 

We further examine the ULySS fits for the cluster members. ULySS computes the redshift difference (hereafter $\Delta cz$) between the redshift derived from the full spectrum and one derived from the best-fit model in the range where we compute the velocity dispersion. There are nine objects with the $\Delta cz$ larger $3 \sigma$ ($\sim 120~\kms$) derived from the $\Delta cz$ distribution for quiescent cluster members. Among these nine objects, four have very large velocity dispersion ($\sigma > 490~\kms$). The ULySS fits to these 4 spectra fail to converge properly because the object shows 1) strong emission lines, 2) an unresolvable night sky subtraction issue, 3) contamination from a nearby star, or 4) an indeterminate redshift as a result of low signal-to-noise. We thus exclude these four objects from further consideration.

We apply an aperture correction to compute consistent velocity dispersions from the SDSS, BOSS, and Hectospec data obtained with different spectrograph apertures. The aperture correction also provides consistent velocity dispersions for cluster galaxies at different redshifts. We use the aperture correction from \citet{2016ApJ...832..203Z}: $\sigma_{A}/\sigma_{B} = (R_{A} / R_{B})^\beta$, where R is the aperture size. The aperture correction coefficient we use is $\beta = -0.059 \pm 0.014$ derived from the galaxies with both Portsmouth and Hectospec velocity dispersion measurements \citep{2019ApJ...872..192S}. This aperture correction is consistent with results from integral field spectroscopy \citep{2006MNRAS.366.1126C}. 
For consistency with previous velocity dispersion functions \citep{2017ApJS..229...20S, 2017ApJ...845...73S, 2019ApJ...872..192S}, we compute the velocity dispersion within a fiducial 3 kpc aperture. The aperture corrected velocity dispersion ($\sigma$) we derive is: $\sigma = \sigma_{raw} * (3 {\rm kpc} / R_{physical}(z))^\beta$, where $\sigma_{raw}$ is the raw velocity dispersion measurement and $R_{physical}(z)$ is the physical size of the fiber aperture of the spectrograph, measured at the redshift of each galaxy. The aperture correction is typically only $\sim 5 \%$. Hereafter, the stellar velocity dispersion ($\sigma$) indicates the value within the 3 kpc aperture. 

\begin{figure*}
\centering
\includegraphics[scale=0.5]{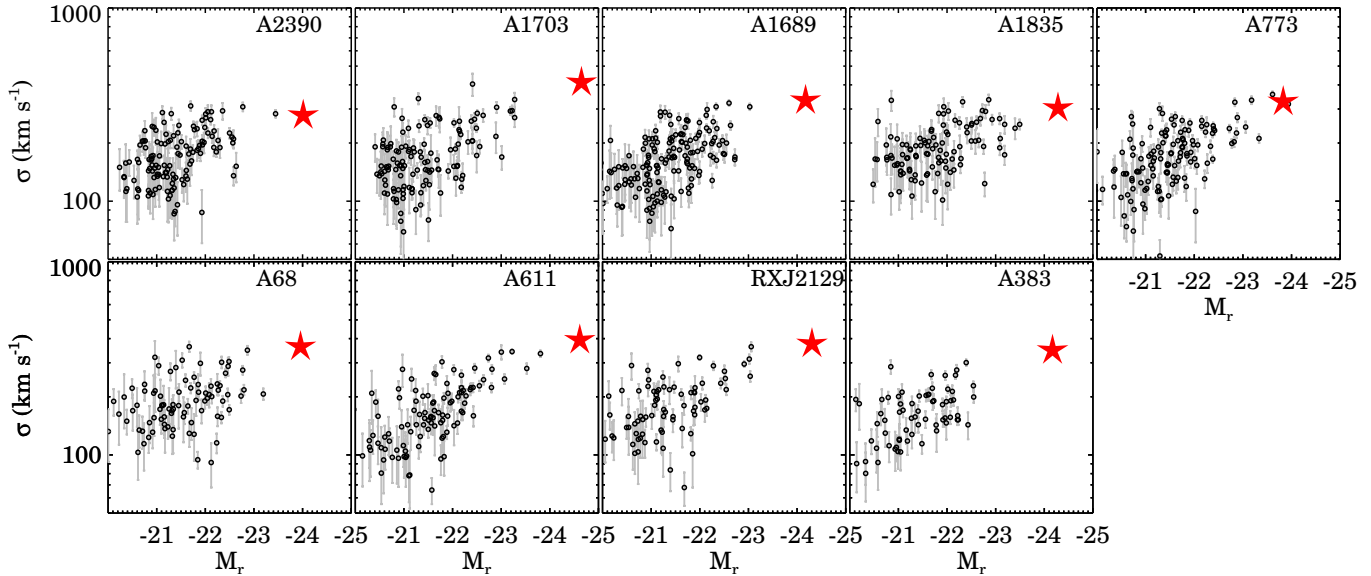}
\caption{Stellar velocity dispersion as a function of $M_{r}$ in each cluster (within $R_{cl} < R_{200}$). Red star symbols indicate the BCG in each cluster. } 
\label{vdisp_mr}
\end{figure*}

Figure \ref{vdisp_mr} shows the stellar velocity dispersion versus the absolute magnitude of members in each cluster. Brighter galaxies generally have larger velocity dispersions. The scatter in the velocity dispersion is larger for less luminous galaxies. This large scatter complicates the conversion between absolute magnitude and velocity dispersion, a relation that is critical for correcting the incompleteness of the spectroscopic survey. 

Red symbols in Figure \ref{vdisp_mr} indicate the brightest cluster galaxies (BCGs) in each cluster. We identify these BCGs among cluster members based on their brightness and we confirm the identification by visual inspection (see \citealp{2020ApJ...891..129S}). Although the BCGs are by definition the brightest member, they do not necessarily have the largest velocity dispersion. Bright galaxies with high velocity dispersions ($M_{r} < -23$) are interesting objects rarely seen in the general field environment \citep{2016ApJ...832..203Z}. There are four out of nine clusters in our sample (A2390, A1835, A773, and A68) that contain galaxies with $\sigma$ larger than that for BCGs. There are 9, 4, 2, and 1 galaxies with $\sigma$ larger than for the BCGs in A2390, A1835, A773, and A68, respectively. The velocity dispersion differences between these galaxies and the BCGs are mostly less than $1\sigma$, except for 3 and 1 galaxies in A1689 and A68 ($< 2 \sigma$), respectively. These galaxies are all red and old (high $\dn$) and their positions are slightly concentrated toward the cluster center. 

%=====================================
% Figure \ref{vdisp_mr}
%=====================================
%=====================================

\begin{figure}
\centering
\includegraphics[scale=0.33]{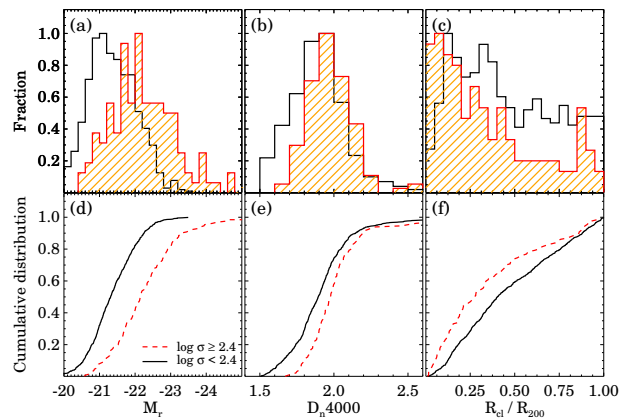}
\caption{Distributions of (a) absolute magnitude, (b) $\dn$, and (c) normalized projected distance from cluster center for cluster members with $\log \sigma < 2.4$ (black open histograms) and with $\log \sigma \geq 2.4$ (red hatched histograms). Panel (d), (e), (f) show the cumulative distributions of $M_{r}$, $\dn$, and normalized clustercentric distance for panels (a)-(c). Black solid and red dashed lines show the cumulative distributions for cluster members with $\log \sigma > 2.4$ and for those with $\log \sigma < 2.4$, respectively} 
\label{excess}
\end{figure}

Figure \ref{excess} contrasts various properties of cluster members with large and small central stellar velocity dispersion. Figure \ref{excess} (a) displays the distributions of $M_{r}$ of cluster members with $\log \sigma > 2.4$ ($\sigma \sim 250~\kms$, hatched histogram) and members with $\log \sigma \leq 2.4$ (open histogram). Figure \ref{excess} (d) shows the cumulative distributions of $M_{r}$ for cluster members with $\log \sigma <2.4$ (black lines) and $\log \sigma \geq 2.4$ (red lines). The two populations show different distributions; the larger $\sigma$ galaxies are generally brighter than their smaller $\sigma$ counterparts. A K-S test also demonstrates that these distributions are not drawn from the same parent distribution ($p = 5.6 \times 10^{-16}$). 

Figure \ref{excess} (b) and (e) compare the distributions of $\dn$ for the two populations. In Figure \ref{excess} (e), the large $\sigma$ population contains more old, high $\dn$ galaxies than the low $\sigma$ population. In other words, the large $\sigma$ galaxies tend to be older than their lower $\sigma$ counterparts. The K-S test shows that indeed the $\dn$ distributions of two populations are not drawn from the same underlying population ($p = 6.0 \times 10^{-12}$). This trend is consistent with the correlation between $\sigma$ and $\dn$ \citep{2017ApJ...841...32Z}. 

Figure \ref{excess} (c) and (f) show the distributions of projected clustercentric distances ($R_{cl} / R_{200}$) of the two populations. The K-S test probability is identical to zero ($p = 6.1 \times 10^{-5}$) indicating that these distributions are not drawn from the same underlying distribution. The large $\sigma$ galaxies show a slight concentration toward cluster center compared to the low $\sigma$ galaxies.

In summary, cluster members with large $\sigma$ are generally brighter, older, and more concentrated toward the cluster center than members with low $\sigma$. These characteristics are similar to the properties of the galaxy populations in Coma and A2029 \citep{2017ApJ...845...73S}. 

%=====================================
% Figure \ref{excess}
%=====================================
%====================================

%============================================================= 
\section{CLUSTER VELOCITY DISPERSION FUNCTIONS}\label{vdf}

The velocity dispersion function counts the number of cluster member galaxies as a function of their velocity dispersion. As in the derivation of the galaxy luminosity, corrections for the inevitable incompleteness of the samples are critical. We introduce the basic idea of the completeness correction in Section \ref{correction}. We describe the empirical corrections we derive based on a larger set of galaxy clusters in Section \ref{hecs}. We derive the cluster velocity dispersion in Section \ref{construction}. In Section \ref{clvdf} we show the resulting cluster velocity dispersion functions. 

\subsection{Completeness Correction for the Velocity Dispersion Function} \label{correction}

The incompleteness of spectroscopic surveys naturally impacts the determination of the velocity dispersion function. Figure \ref{zcomp_mag} indicates that the spectroscopic completeness of our survey falls to $50\%$ at $\sim 20.5$ for each cluster. At the median $z = 0.23$ for the cluster sample, $r =20.5$ corresponds to $M_{r} = -19.8$, corresponding roughly to a velocity dispersion $\sigma = 100~\kms$ (see Figure \ref{vdisp_mr}). The incompleteness correction and its uncertainty are substantial for velocity dispersions $\lesssim 160~\kms$. 

We also correct for cluster members with spectra of insufficient quality to yield a velocity dispersion. The dashed lines in Figure \ref{zcomp_mag} show the velocity dispersion completeness within $R_{200}$ for each cluster. Fortunately, the correction for galaxies with a spectrum but without a velocity dispersion is small in our sample clusters because we measure the velocity dispersion for most cluster members.

The overall completeness correction requires a two-step process: 1) estimation of the number of missing cluster members in the photometric sample and 2) estimation of the velocity dispersions of a missing cluster member from its luminosity. To count the number of missing cluster members, the member fraction among photometric galaxies is required. The derivation of the member fraction is complicated because the member fraction is a function of absolute magnitude ($M_{r}$), clustercentric distance ($R_{cl}$), and $\dn$ (or color). We use the member fraction measured in each sample cluster (Section \ref{construction}). 

Furthermore, the estimation of the velocity dispersion from the galaxy luminosity is also not straightforward. Figure \ref{vdisp_mr} (see also \citealp{2016ApJ...832..203Z, 2017ApJ...845...73S}) shows the large scatter around the correlation between velocity dispersion and luminosity. This correlation between velocity dispersion and luminosity (or stellar mass) also depends on $\dn$ \citep{2017ApJ...841...32Z}. To reduce the uncertainty in the velocity dispersion estimated from luminosity, we derive an empirical relation based on a large sample of cluster galaxies with luminosity, $\dn$, and velocity dispersion measurements.

\subsection{Empirical Distributions from HeCS-omnibus} \label{hecs}

The HeCS-omnibus dataset \citep{2020ApJ...891..129S} provides the empirical distributions that are the basis for incompleteness correction. HeCS-omnibus is a compilation of spectroscopic data for 225 galaxy clusters at $0.02 < z < 0.29$ \citep{2006AJ....132.1275R, 2013ApJ...767...15R, 2016ApJ...819...63R, 2018ApJ...862..172R, 2015ApJ...806..101H, 2017ApJS..229...20S}. HeCS-omnibus includes cluster membership based on the caustic technique. K-corrected absolute magnitudes and central velocity dispersions of the HeCS-omnibus galaxies are analogous to the quantities measured here. To avoid introducing systematic bias, we select 49 well observed clusters within the redshift range of our target clusters ($0.18 < z < 0.29$) as a foundation for the correction. 

We derive the velocity dispersion distribution of the HeCS-omnibus galaxies in various $M_{r}$ and $\dn$ bins (the luminosity-$\sigma$-$\dn$ relations). \citet{2017ApJS..229...20S} derived a similar probability distribution functions based on SDSS field galaxies and used them for estimating missing velocity dispersions. The relationship between absolute magnitude and central velocity dispersion is essentially identical for the HeCS-omnibus clusters and the sample of nine clusters we describe in detail here.

\begin{figure*}
\centering
\includegraphics[scale=0.53]{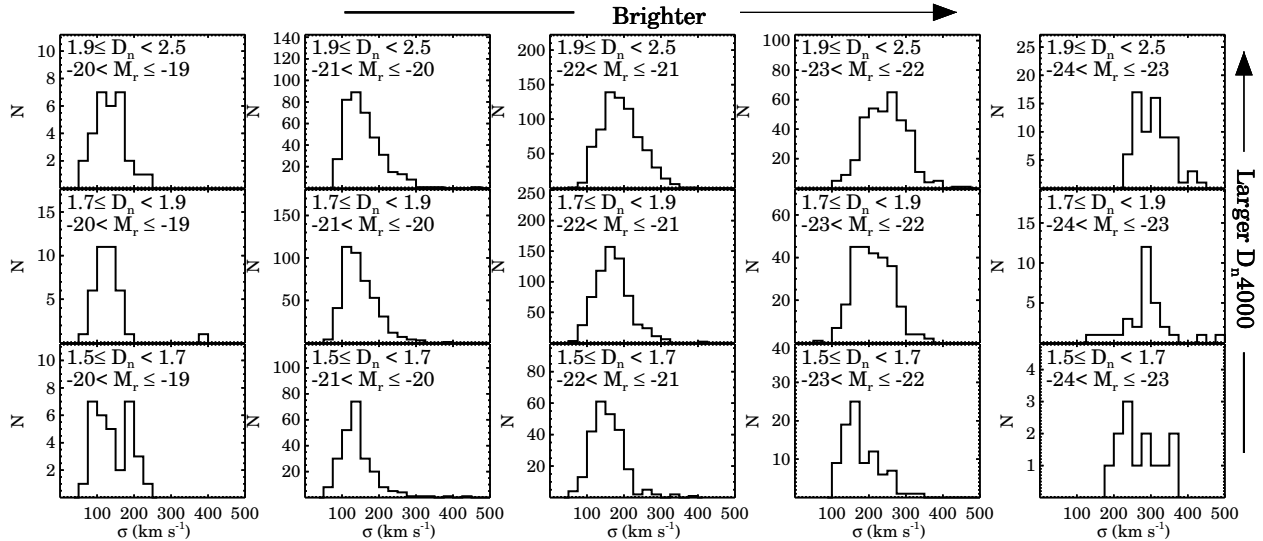}
\caption{
Velocity dispersion distributions of the HeCS-omnibus cluster members in various $\dn$ and absolute magnitude bins. Panels on the right side show brighter cluster members and the upper panels show cluster members with larger $\dn$. } 
\label{sigma_dn_hist}
\end{figure*}

Figure \ref{sigma_dn_hist} shows the velocity dispersion distribution of the HeCS omnibus galaxies within $R_{200}$ in various $M_{r}$ and $\dn$ bins. In each cluster, we compute $M_{r}$ for all photometric galaxies within $R_{200}$ by assuming they have the mean cluster redshift. We then stack all HeCS-omnibus galaxies in $M_{r}$ and $R_{cl}$ bins; $-19 > M_{r} \geq -24$ in 1.0 magnitude bins and $R_{cl} / R_{200}$ in 0.25 $R_{cl} / R_{200}$ bins. From the stacked sample, we derive the velocity dispersion distributions in three $\dn$ bins ($1.5 \leq \dn < 1.7$, $1.7 \leq \dn < 1.9$. and $1.9 \leq \dn$). Based on these velocity dispersion distributions, we assign the velocity dispersion for galaxies without a measurement based on their $M_{r}$ and $\dn$. 

%=====================================
% Figure \ref{sigma_dn_hist}
%=====================================
%=====================================

\subsection{Construction of the Velocity Dispersion Function} \label{construction}

\begin{figure*}[h!]
\centering
\includegraphics[scale=0.7]{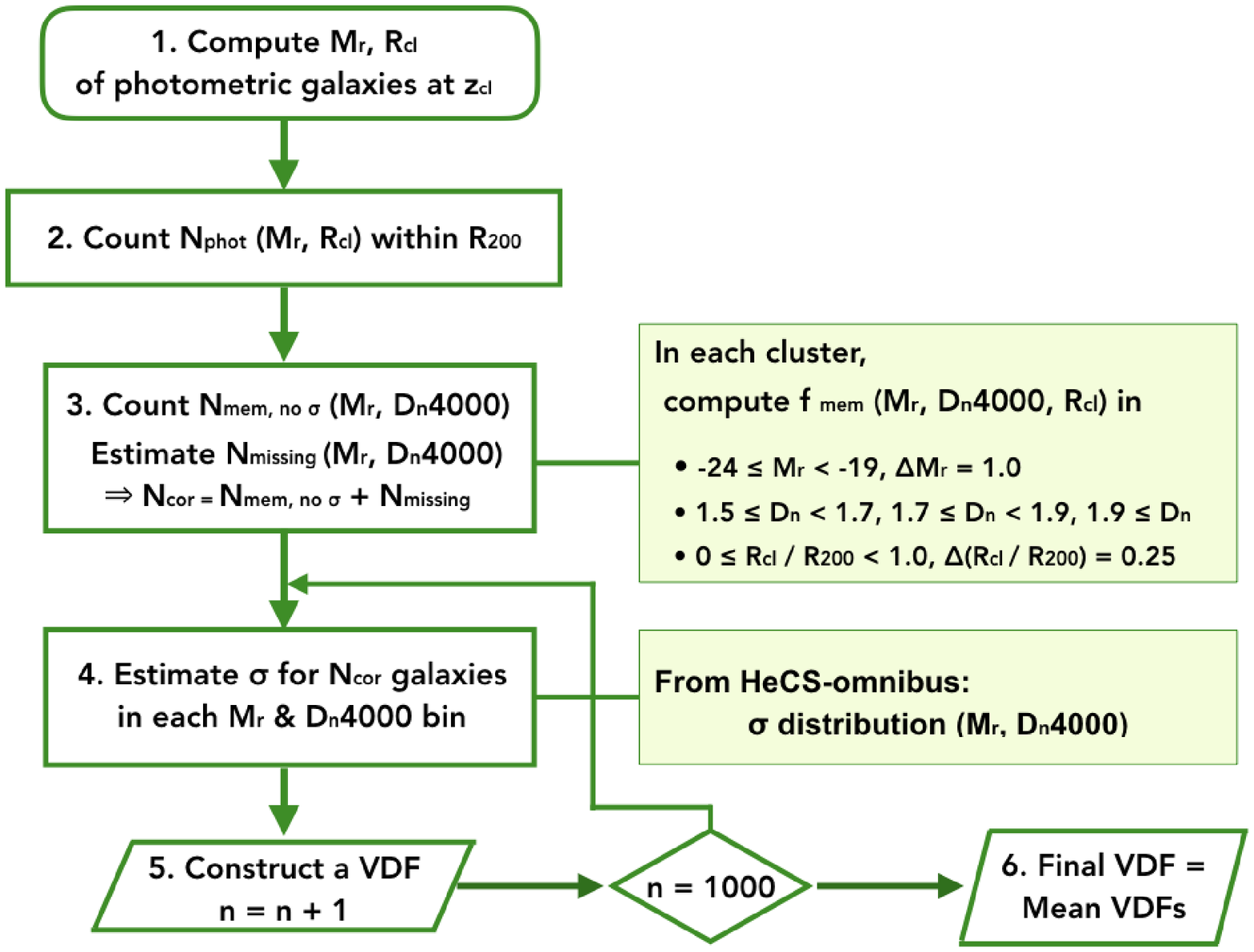}
\caption{Flow chart summarizing the derivation of the cluster velocity dispersion functions. } 
\label{flow}
\end{figure*}

We correct the velocity dispersion function following the approach in \citet{2017ApJS..229...20S}. Figure \ref{flow} summarizes the correction process. The steps in Figure \ref{flow} are: 
\begin{enumerate}
\item{We compute the absolute magnitudes of galaxies in the photometric sample by assuming that each photometric galaxy without a spectrum is at the mean cluster redshift. } 

\item{We count the number of photometric galaxies in $M_{r}$ and $R_{cl}/R_{200}$ bins ($N_{\rm phot} (M_{r}, R_{cl}/R_{200})$). The bins in $M_{r}$ and $R_{cl}/R_{200}$ are the same as the ones we use for the spectroscopic member fraction for HeCS-omnibus (Figure \ref{sigma_dn_hist}).}

\item{We estimate the number of member galaxies ($N_{cor}$) that need a statistical $\sigma$ estimate (galaxies without a $\sigma$ measurement) in $M_{r}$ and $\dn$ bins. Here, $N_{cor}$ is defined as
\begin{align*}
 N_{cor} (M_{r}, \dn) &= N_{mem, no~ \sigma} (M_{r}, \dn) \\
 &+ N_{missing} (M_{r}, \dn),   
\end{align*} where $N_{mem, no~\sigma}$ is the number of spectroscopic members without a $\sigma$ measurement and $N_{missing} (M_{r}, \dn)$ is the number of statistically probable member galaxies without spectra. We estimate the number of probable member galaxies ($N_{missing}$) from the number of photometric galaxies without spectra in $M_{r}, R_{cl}$ and $\dn$ bins: 
\begin{align*}
N_{missing} &(M_{r}, \dn) = \\
    &\sum \limits_{R' = 0}^1 (N_{phot} (M_{r}, R') - N_{spec} (M_{r}, R')) \\ 
    &\times f_{mem} (M_{r}, R', \dn), 
\end{align*}
where $R'$ is $R_{cl} / R_{200}$, $N_{phot} (M_{r}, R')$ and $N_{spec} (M_{r}, R')$ indicate the number of photometric and spectroscopic galaxies in each $M_{r}$ and $R_{cl}/R_{200}$ bin. The spectroscopic member fraction, $f_{mem} (M_{r}, R', \dn)$ is defined as follows: 
\begin{align*}
f_{mem} (M_{r}, R', &\dn) = \frac{N_{spec. mem} (M_{r}, R')}{N_{phot} (M_{r}, R')}\\
&\times \frac{N_{spec. mem} (M_{r}, R', \dn)}{N_{spec. mem} (M_{r}, R')}. 
\end{align*}
Here, $N_{spec} (M_{r}, R', \dn)$ is the number of spectroscopic members in each $M_{r}$, $R_{cl}/R_{200}$ and $\dn$ bin. We calculate the member fraction in $\dn$ bins, because we have to estimate the $\dn$ of missing member galaxies in order to estimate their velocity dispersion. Estimating the $\dn$ of missing member galaxies determines the luminosity - $\sigma$ - $\dn$ relation (see Figure \ref{sigma_dn_hist}) we use to compute $\sigma$s. We note that our analysis is robust to bin size changes. }

\item{We randomly draw $N_{cor} (M_{r}, \dn)$ velocity dispersions from the parent velocity dispersion distribution determined from the HeCS-omnibus sample (Figure \ref{sigma_dn_hist}). } 

\item{The number of cluster members in each velocity dispersion bin is the velocity dispersion function. }

\item{We repeat step 4 and 5 a thousand times each and take the mean as the final estimate of the velocity dispersion function. We propagate the uncertainties in the number counts for missing galaxies to determine the uncertainty in the velocity dispersion function. }
\end{enumerate} 

%=====================================
% Figure \ref{flow}
%=====================================
%===================================== 

\subsection{Cluster Velocity Dispersion Functions} \label{clvdf}

\begin{figure*}[h!]
\centering
\includegraphics[scale=0.47]{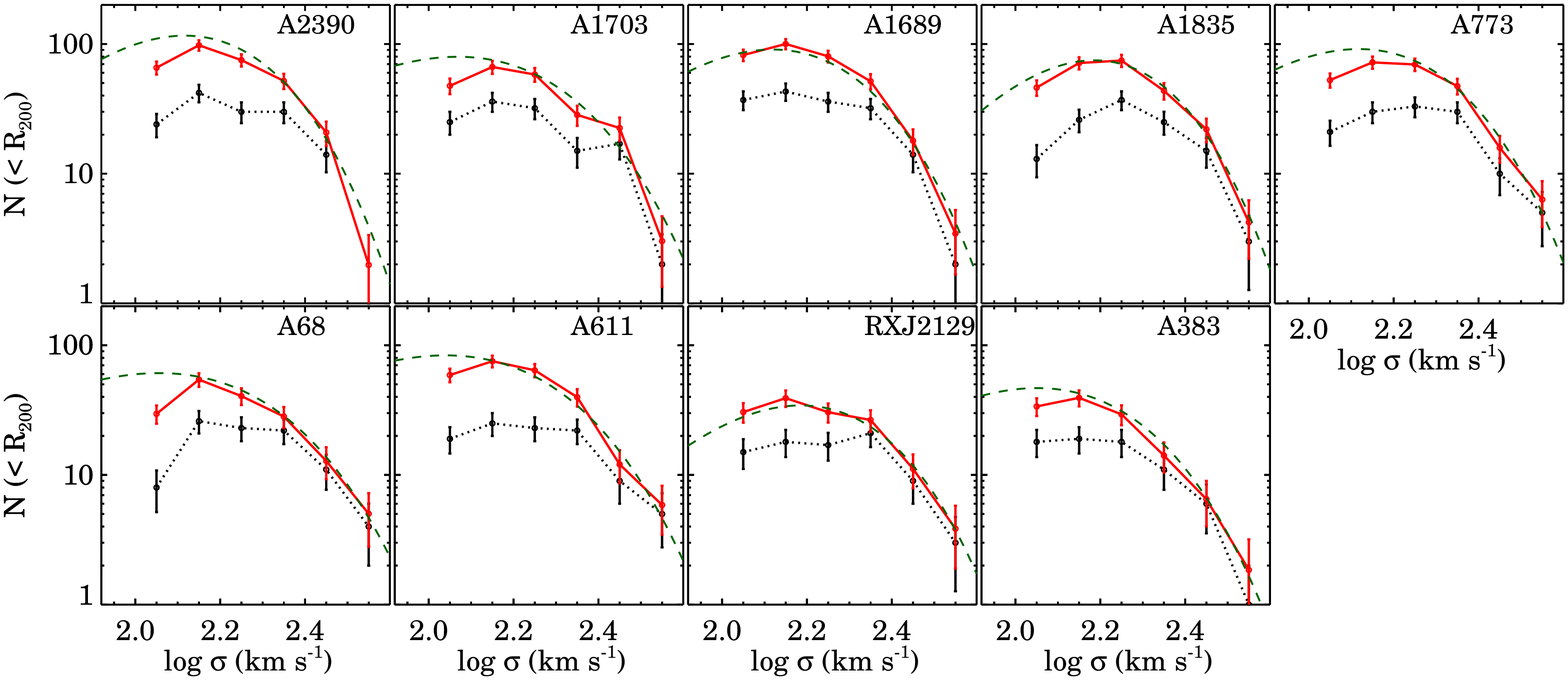}
\caption{The velocity dispersion functions. We use the cluster members within $R_{cl} < R_{200}$. Black symbols show the raw velocity dispersion function and red symbols indicate the corrected velocity dispersion function. Green dashed-lines are the best-fit Schechter functions. } 
\label{vdf_individual}
\end{figure*}

Figure \ref{vdf_individual} shows the individual cluster velocity dispersion functions (VDFs) sorted by dynamical mass ($M_{200}$). Black and red symbols show the raw and corrected VDFs, respectively. The correction for missing velocity dispersions is larger at low velocity dispersion. There is a small correction at large velocity dispersion ($\log \sigma > 2.3$). This small correction results from the large spread in the velocity dispersion - magnitude relation for fainter galaxies; faint objects with large central velocity dispersion account for the small correction at large dispersion. Table \ref{vdf_tab} contains the corrected VDF for each sample cluster. 

%=================================
%Table \ref{list}
%=================================
\begin{deluxetable*}{lccccccccc}
\tablenum{3}
\tablecaption{Corrected Velocity Dispersion Functions \label{vdf_tab}}
\tablecolumns{10}
\tabletypesize{\scriptsize}
\tablewidth{0pt}
\tablehead{\colhead{$\log \sigma$} & \colhead{A2390} & \colhead{A1703} & \colhead{A1689} & \colhead{A1835} & \colhead{A773} & \colhead{A68} & \colhead{A611} & \colhead{RXJ2129} & \colhead{A383}} 
\startdata
2.05 & $65.60 \pm 7.64$ & $47.60 \pm 6.37$ & $ 81.61 \pm 8.36$ & $45.52 \pm 6.22$ & $52.50 \pm 6.66$ & $29.59 \pm 4.92$ & $58.66 \pm 6.96$ & $31.02 \pm 5.21$ & $34.08 \pm 5.30$ \\
2.15 & $97.71 \pm 9.10$ & $66.71 \pm 7.54$ & $100.10 \pm 9.12$ & $71.63 \pm 7.48$ & $72.54 \pm 7.72$ & $54.12 \pm 6.64$ & $75.46 \pm 7.69$ & $38.97 \pm 5.65$ & $39.59 \pm 5.47$ \\
2.25 & $75.20 \pm 7.93$ & $58.18 \pm 6.96$ & $ 81.61 \pm 8.33$ & $74.86 \pm 7.94$ & $69.35 \pm 7.64$ & $40.61 \pm 6.00$ & $64.37 \pm 7.39$ & $30.48 \pm 5.08$ & $29.12 \pm 5.04$ \\
2.35 & $51.58 \pm 6.90$ & $28.65 \pm 4.96$ & $ 51.40 \pm 6.84$ & $43.76 \pm 6.20$ & $47.32 \pm 6.61$ & $28.15 \pm 5.22$ & $39.64 \pm 6.14$ & $26.58 \pm 4.99$ & $14.11 \pm 3.64$ \\
2.45 & $20.63 \pm 4.43$ & $22.57 \pm 4.52$ & $ 17.92 \pm 4.06$ & $21.87 \pm 4.51$ & $15.81 \pm 3.72$ & $12.91 \pm 3.55$ & $12.26 \pm 3.39$ & $11.12 \pm 3.24$ & $ 6.49 \pm 2.48$ \\
2.55 & $ 1.94 \pm 1.39$ & $ 2.94 \pm 1.68$ & $ 3.49 \pm 1.76$ & $  4.22 \pm 2.01$ & $ 6.35 \pm 2.42$ & $ 4.96 \pm 2.23$ & $ 5.86 \pm 2.43$ & $ 3.76 \pm 1.95$ & $ 1.83 \pm 1.35$ \\
\enddata
\end{deluxetable*}
%=================================

%=====================================
% Figure \ref{vdf_individual}
%=====================================
%=====================================

We fit the cluster VDFs with a Schechter function in the range $2.1 < \log \sigma < 2.6$ (green dashed lines in Figure \ref{vdf}). The Schechter function is widely used to characterize luminosity or stellar mass functions. Previous VDF studies sometimes use a modified Schechter function (equation 4 in \citealp{2003ApJ...594..225S}) to fit the VDF \citep{2007ApJ...658..884C, 2010MNRAS.402.2031C, 2016MNRAS.461.1131M, 2017ApJS..229...20S, 2017ApJ...845...73S}. We do not follow this practice because the modified Schechter function has an additional parameter that is poorly constrained with the present data. We summarize the best-fit Schechter function parameters in Table \ref{fit}. 

%=================================
%Table \ref{list}
%=================================
\begin{deluxetable}{lcc}
\tablecaption{Best-Fit Schechter Function Parameters \label{fit}}
\tablecolumns{3}
\tabletypesize{\scriptsize}
\tablewidth{0pt}
\tablehead{
\colhead{ID} & \colhead{$\alpha$} & \colhead{$\sigma_{*}$ (km s$^{-1}$)}}
\startdata
  A2390 & $4.40 \pm 1.08$ & $28.93^{+ 5.48}_{- 4.61}$ \\ 
  A1703 & $2.66 \pm 1.13$ & $40.02^{+11.44}_{- 8.90}$ \\ 
  A1689 & $3.56 \pm 1.09$ & $33.67^{+ 7.59}_{- 6.20}$ \\ 
  A1835 & $4.08 \pm 1.09$ & $32.72^{+ 7.13}_{- 5.86}$ \\ 
   A773 & $3.22 \pm 1.15$ & $38.25^{+10.43}_{- 8.19}$ \\ 
    A68 & $3.52 \pm 1.71$ & $36.09^{+15.53}_{-10.86}$ \\ 
   A611 & $2.32 \pm 1.26$ & $45.44^{+16.76}_{-12.24}$ \\ 
RXJ2129 & $1.75 \pm 1.35$ & $53.66^{+25.53}_{-17.30}$ \\ 
   A383 & $2.27 \pm 1.72$ & $42.02^{+21.00}_{-14.00}$ \\ 
\hline     
   Coma\tablenotemark{*} & $4.90 \pm 1.80$ & $24.75^{+ 7.42}_{-5.71}$ \\
  A2029\tablenotemark{*} & $4.46 \pm 1.70$ & $27.85^{+ 8.88}_{-6.73}$
\enddata
\tablenotetext{*}{VDFs from \citet{2017ApJS..229...20S}. We derive the best Schechter function fit here.}
\end{deluxetable}
%=================================

\section{COMPARISON WITH OTHER VELOCITY DISPERSION FUNCTIONS}\label{comparison}

\subsection{Comparison with Local Massive Clusters}\label{comp_cl}

There are measurements of galaxy velocity dispersion functions for only a few clusters. \citet{2016ApJ...827L...5M} examine the circular velocity distribution of a nearby cluster, A2142. \citet{2017ApJS..229...20S} measured the VDFs of the two local massive clusters, Coma and A2029, with $\sim 500$ quiescent galaxy velocity dispersion measurements in each cluster. The shapes of Coma and A2029 VDFs are essentially identical. 

Here we compare the cluster VDFs for the current sample with the Coma and A2029 VDFs. This comparison is straightforward because all the VDFs are measured consistently. \citet{2017ApJS..229...20S} also use 3 kpc aperture corrected velocity dispersions for quiescent galaxies with $\dn > 1.5$. They also correct empirically for incompleteness in the VDFs using a similar technique.

%=====================================
% Figure \ref{vdf_cl}
%=====================================
\begin{figure}[ht!]
\centering
\includegraphics[scale=0.5]{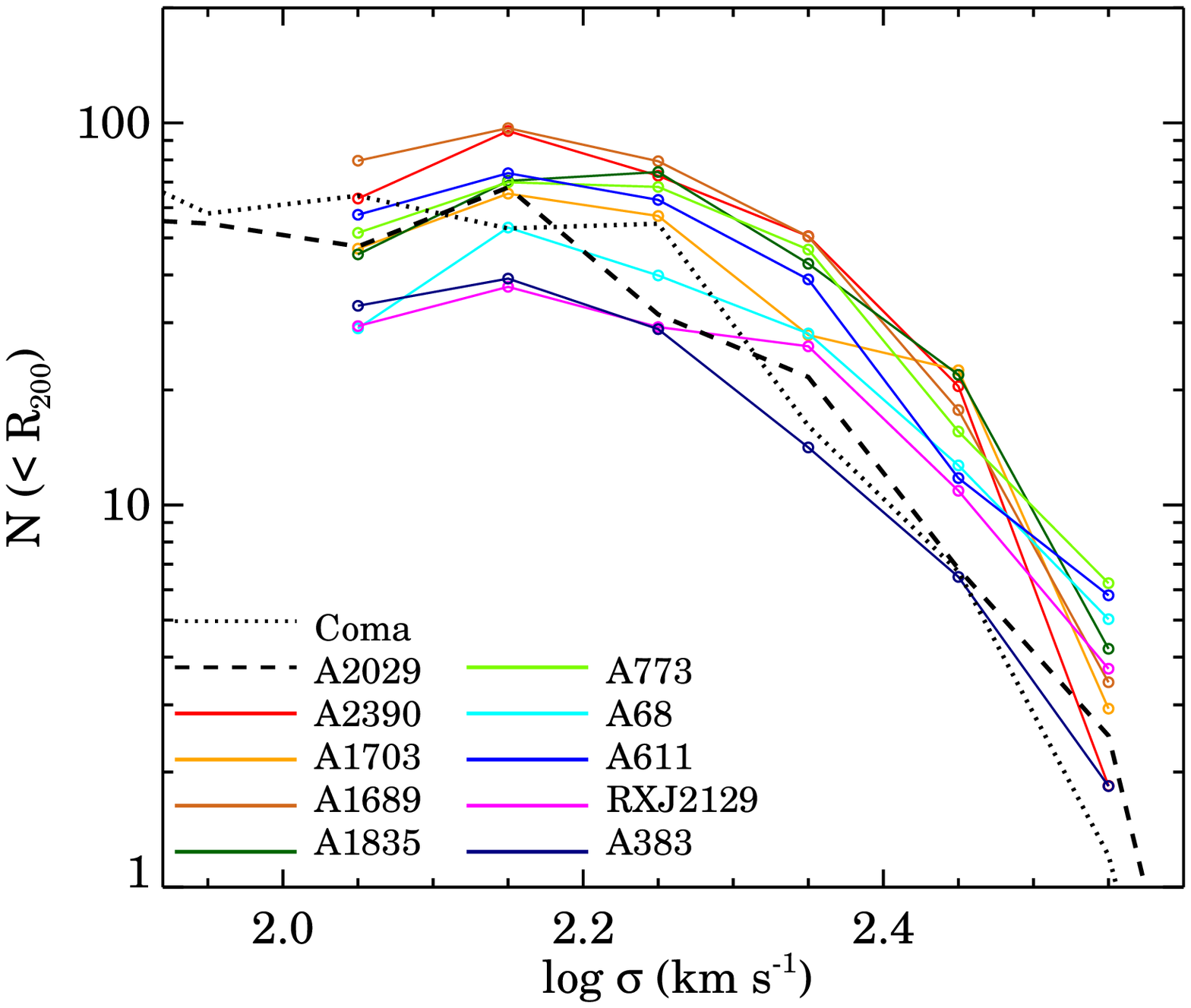}
\caption{Cluster velocity dispersion functions compared to Coma and A2029. }
\label{vdf_cl}
\end{figure}
%=====================================

Figure \ref{vdf_cl} displays VDFs for the current sample (solid lines) and for Coma (dotted line) and A2029 (dashed line). The cluster VDFs generally have similar shapes although the normalizations are slightly different. The normalization depends on cluster mass; more massive clusters tend to contain more galaxies. 

\begin{figure*}
\centering
\includegraphics[scale=0.67]{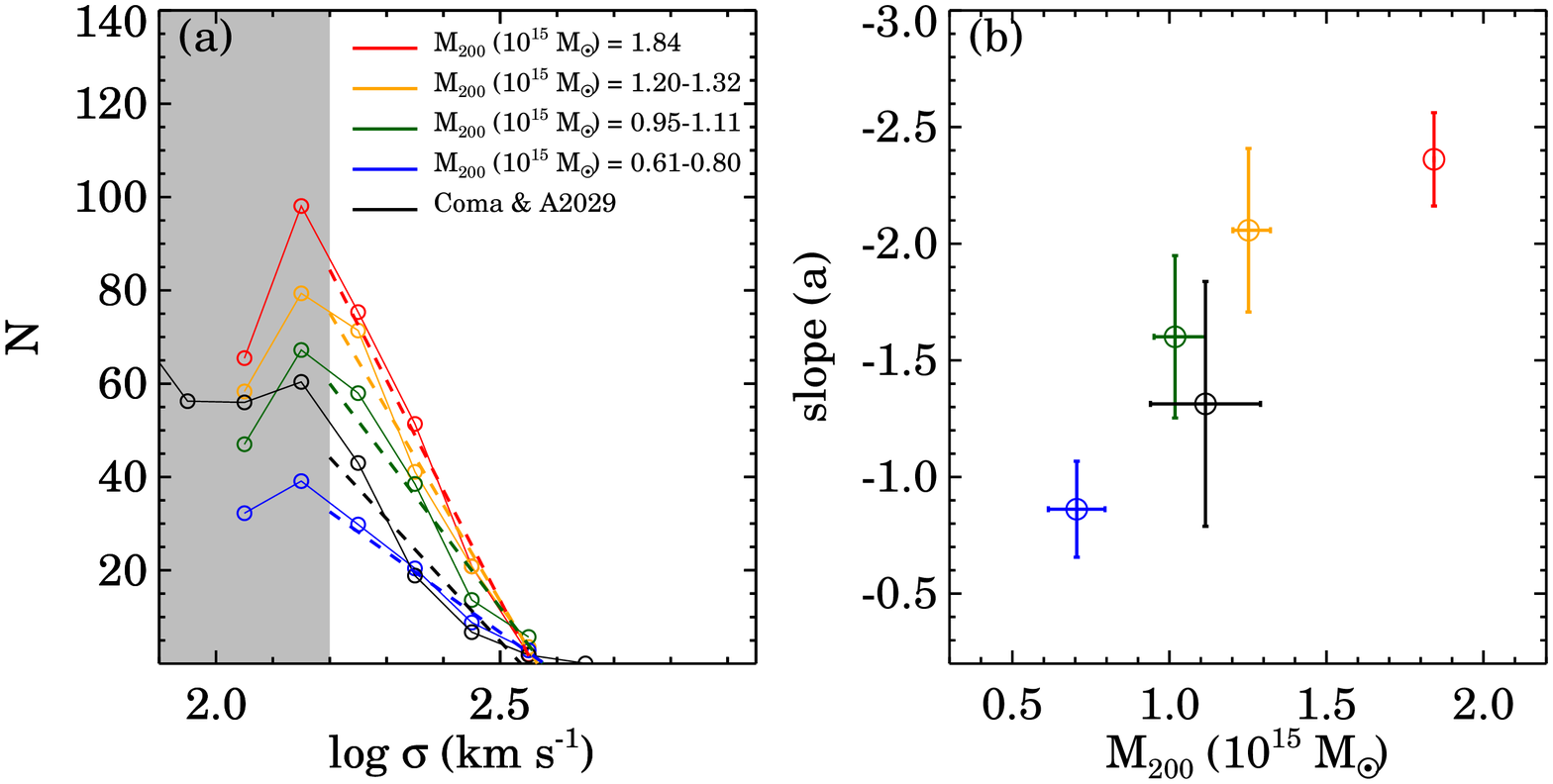}
\caption{(a) Comparison of cluster velocity dispersion functions (solid lines) binned in cluster mass. The dotted lines show the best-fit power-law to the cluster VDFs at $2.2 < \log \sigma < 2.6$ (non-shaded region). (b) High $\sigma$ slope of the cluster VDFs as a function of cluster mass. } 
\label{slope}
\end{figure*}

Figure \ref{slope} (a) allows direct comparison of the VDF slopes at $\log \sigma > 2.2$. We derive the mean cluster VDFs for the clusters in four mass bins: $0.61 \leq (M_{200} / 10^{15} M_{\odot}) \leq 0.80$, $0.95 \leq (M_{200} / 10^{15} M_{\odot}) \leq 1.11$, $1.20 \leq (M_{200} / 10^{15} M_{\odot}) \leq 1.32$, and $(M_{200} / 10^{15} M_{\odot}) = 1.84$. We normalized cluster VDFs at $\log \sigma = 2.25$ to compare the shape of the VDFs at the large velocity dispersion end. 

The VDFs steepen with increasing cluster mass; the low mass clusters have a larger proportion of high $\sigma$ galaxies. For example, A773 and A68 have $\sim 6$ galaxies with $\log \sigma > 2.5$; the most massive cluster in our sample, A2390, has only two galaxies with $\log \sigma > 2.5$. We derive the best-fit power law at $\log \sigma > 2.2$: $(N/100) = a \log_{10} \sigma + b$.  Figure \ref{slope} (b) plots the power-law slope ($a$) as a function of cluster mass. Although the slope measurement uncertainties are large, the slope clearly increases with cluster mass. 

The slope of the cluster VDF is determined mainly by the number of galaxies in the largest $\sigma$ bin ($2.5 < \log \sigma \leq 2.6$). In other words, more massive clusters include fewer galaxies in the largest $\sigma$ bin. The interpretation of this result is subtle. Massive clusters indeed have large $\sigma$ BCGs, and the $\sigma$ of BCG is proportional to the cluster mass \citep{2020ApJ...891..129S}. In other words, more massive clusters have fewer galaxies with $2.5 < \log \sigma \leq 2.6$, but their members in this bin have larger $\sigma$.

%=====================================
% Figure \ref{vdf_individual}
%=====================================
%=====================================

\subsection{Comparison with the Field Velocity Dispersion Function}\label{comp_field}

There are several VDF measurements for quiescent (or early-type) galaxies in the field based on SDSS and/or BOSS spectroscopy \citep{2003ApJ...594..225S, 2005ApJ...622...81M, 2007ApJ...658..884C, 2010MNRAS.402.2031C, 2017MNRAS.468...47M, 2017ApJ...845...73S}. These field VDFs differ from one another possibly due to differences in sample selection.

\citet{2017ApJ...845...73S} compare the field VDF with the cluster VDFs for Coma and A2029. To derive a complete field VDF, they construct a $\sigma$-complete sample at $0.03 \leq z \leq 0.10$ from SDSS spectroscopy. This $\sigma$-complete sample contains a complete set of quiescent galaxies with $\sigma$ larger than the $\sigma$ measurement limit of SDSS. The use of $\sigma$-complete sample significantly reduces systematic bias in the field VDF measurements. They show that the field and the cluster (Coma and A2029) VDFs are consistent at $\log \sigma < 2.4~(\sigma \sim 250~\kms)$, but the cluster VDFs show a significant excess at $\log \sigma > 2.4$. 

We compare the cluster VDFs with the field VDF from \citet{2017ApJ...845...73S}. We use the same definition of $\sigma$ and quiescent galaxy selection ($\dn > 1.5$) as used in \citet{2017ApJ...845...73S}. We scale the field VDF and cluster VDFs to allow a clearer comparison of the VDF shapes. 

\begin{figure}
\centering
\includegraphics[scale=0.49]{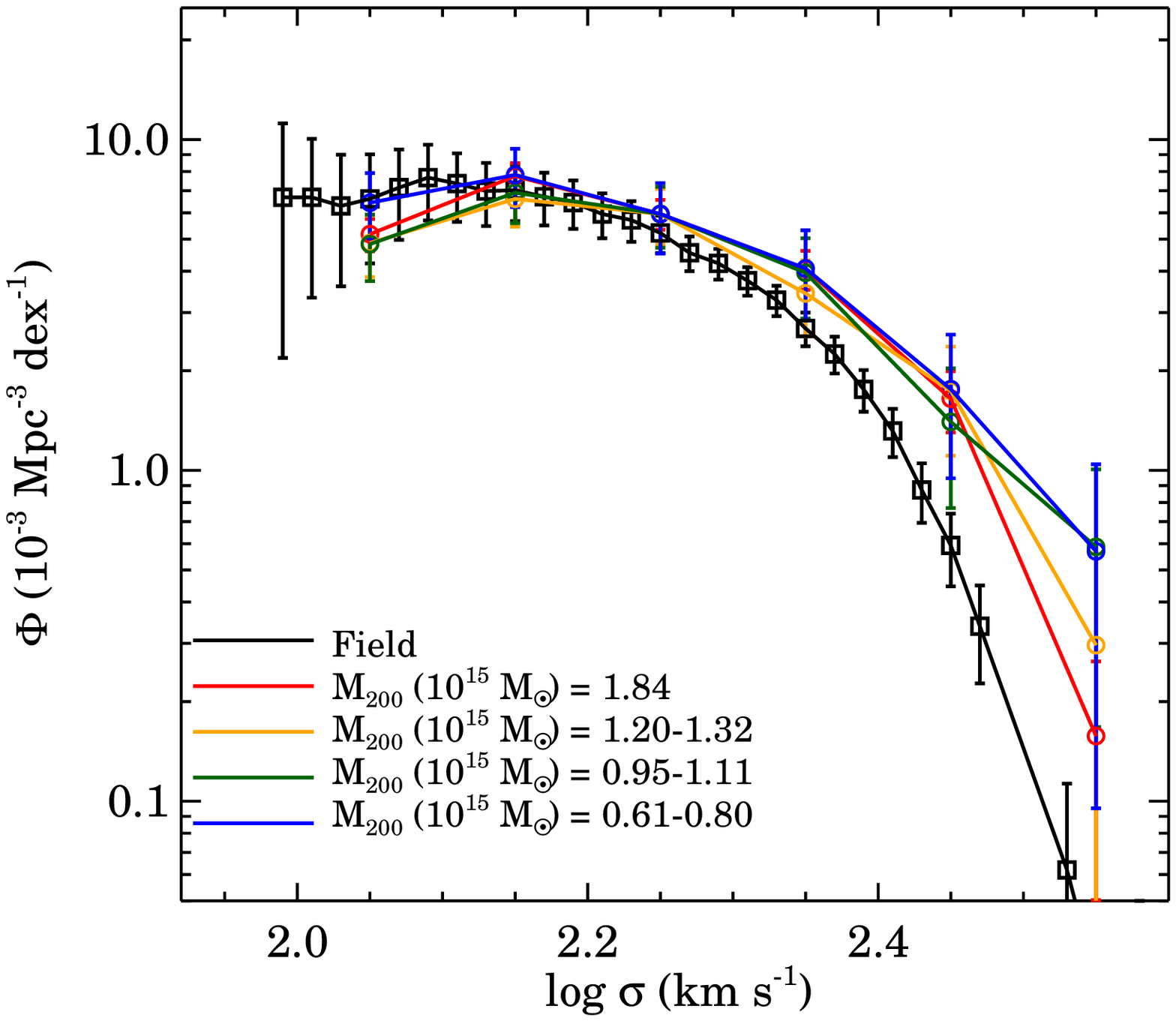}
\caption{Velocity dispersion function for quiescent galaxies in the local universe ($0.03 < z < 0.10$, \citealp{2017ApJ...845...73S}) compared to cluster VDFs in four mass bins. }
\label{field}
\end{figure}

Figure \ref{field} compares the field and cluster VDFs. We plot the cluster VDFs averaged in $M_{200}$ bins. The cluster VDFs show an excess at $\log \sigma > 2.4$ compared to the field VDF regardless of the cluster mass. Coma and A2029 VDFs show a similar excess relative to the field \citep{2017ApJS..229...20S}. The excess results from the presence of multiple galaxies usually including the BCG with high $\sigma$ in each cluster. This excess in clusters relative to the field population is also present in the galaxy luminosity functions \citep{2003ApJ...591..764C, 2004ApJ...617..879L, 2012A&A...540A..90B}. Large $\sigma$ (or high mass) galaxies, whether or not they are the BCG, are relatively more abundant in clusters than in the field. 

The contrast between the field and cluster VDFs along with the changing slope of the VDF as a function of cluster mass present a conundrum. At group masses, the excess at high $\sigma$ must disappear and the slope must increase to match the field result. Groups with masses of $\sim 10^{13-13.5}$ are abundant and thus make a major contribution to the field VDF. 

%=====================================
% Figure \ref{field}
%=====================================
%=============================================================

%=============================================================
\section{Origins of the Cluster Velocity Dispersion Function Excess} \label{discussion}

The cluster velocity dispersion functions we derive reveal two interesting features. All of the cluster VDFs exceed the field VDF at $\log \sigma \geq 2.4$; this excess is consistent with the excess seen in local massive cluster VDFs \citep{2017ApJ...845...73S}. We also show that the slope of the cluster VDF at $\log \sigma > 2.2$ depends on cluster mass. We explore the physical properties of large $\sigma$ galaxies in Section \ref{lvdisp}. In Section \ref{theory}, we discuss current theoretical models for the origin of the cluster VDFs. We discuss observational limitations and the interpretation of the cluster VDFs in Section \ref{limit}. 

\subsection{Spatial Distribution of Large $\sigma$ Galaxies} \label{lvdisp}

The cluster VDFs exceed the field VDF for $\log \sigma \geq 2.4$. These large $\sigma$ galaxies are generally brighter, older galaxies, and they are more concentrated towards the cluster center than cluster members with lower $\sigma$. 

\begin{figure*}
\centering
\includegraphics[scale=0.47]{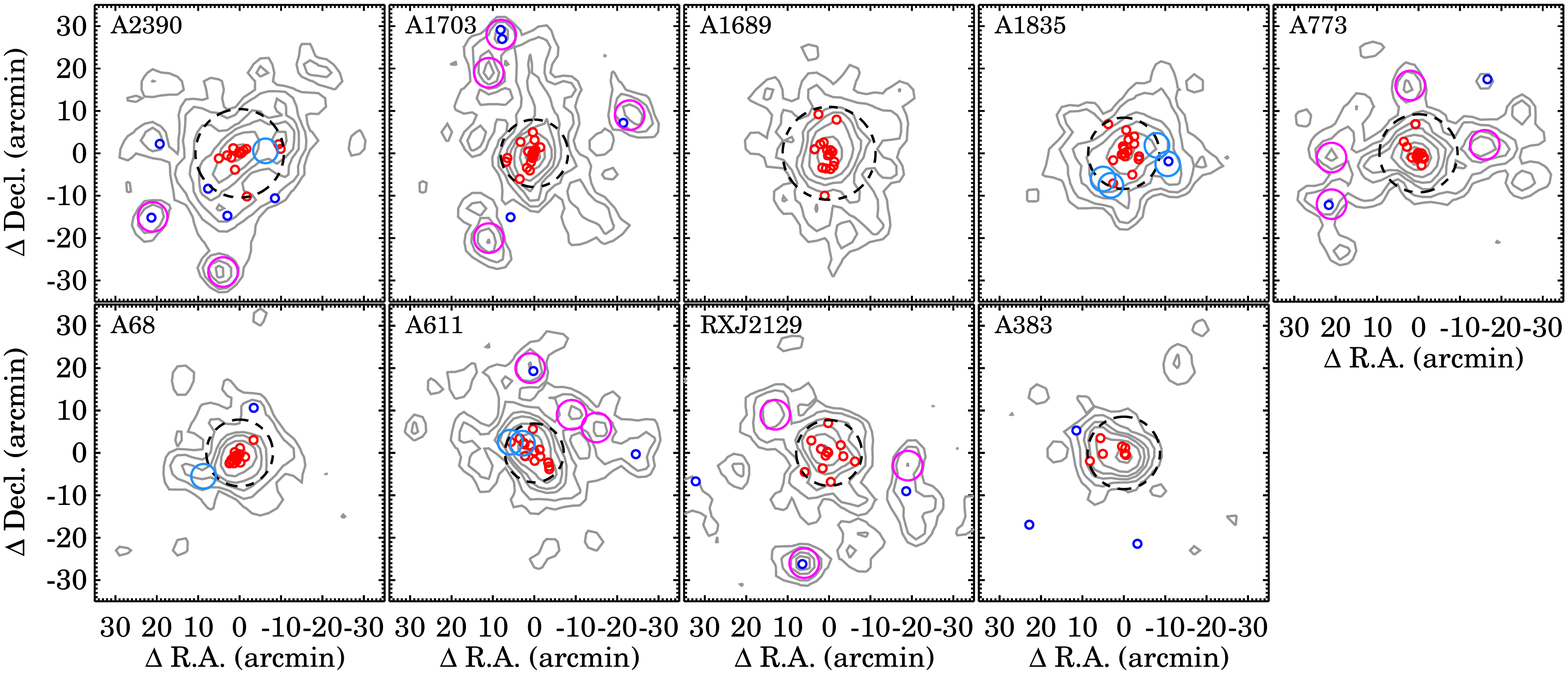}
\caption{
Number density maps of spectroscopically identified members in each cluster. Black dashed circles show $R_{200}$. Red and blue circles mark the locations of galaxies $\sigma \geq 250$ km s$^{-1}$ within $R_{200}$ and outside $R_{200}$, respectively. Large magenta circles indicate the substructures we identify based on the number density map of spectroscopic members. Large blue circles indicate the X-ray subsystems identified in \citet{2018MNRAS.477.4931H}. } 
\label{spatial}
\end{figure*}

The detailed spatial distribution of the cluster members with $\log \sigma > 2.4$ including those at outside $R_{200}$ provides clues to the origin of the VDF excess. Figure \ref{spatial} shows number density maps (contours) for the spectroscopic members in each cluster. Red circles mark the location of large $\sigma$ galaxies with $\log \sigma > 2.4$ that contribute to the excess in the cluster VDFs. These large $\sigma$ galaxies are concentrated towards the cluster center, consistent with Figure \ref{excess} (c). We also show the location of the cluster members with $\log \sigma > 2.4$ outside $R_{200}$ with blue symbols. 

%=====================================
% Figure \ref{spatial}
%=====================================
%=====================================

The number density maps of some clusters show local density peaks (magenta circles in Figure \ref{spatial}). These local density peaks are typically $\sim 200$ times denser than the mean number density of spectroscopic members at the same clustercentric distance. These local density peaks often correspond to cluster substructures. For example, \citet{2019ApJ...871..129S} explore the spatial distribution of A2029 members (their Figure 5); there are two dense substructures in the infall region ($R_{cl}/R_{200} \sim 1.3 - 1.6$). These substructures also appear in both X-ray and weak lensing maps. \citet{2019ApJ...872..192S} demonstrate that the infalling substructures will be accreted onto the main cluster within 2-3 Gyr timescale. Similarly, the local density peaks we identify in the number density maps of cluster galaxies are potential infalling substructures because they are contained within the caustics. 

We check the X-ray visibility of the local density peaks in the member density maps. X-ray images of these clusters from the ROSAT all-sky survey are too shallow to identify X-ray substructures. \citet{2007A&A...467..485H} use {\it Chandra} images to investigate the X-ray properties of six clusters in our sample (see also \citealp{2014MNRAS.440..588H}). However, the substructures we identify are not covered by the {\it Chandra} images. There are {\it XMM} images of all of our target clusters. Similar to {\it Chandra} images, the {\it XMM} images do not cover the region where we identify local peaks. 

\citet{2018MNRAS.477.4931H} use {\it XMM} X-ray images to identify eight infalling groups in four clusters in our sample (A2390, A1835, A68, A611; large blue circles in Figure \ref{spatial}). Because these X-ray groups are close to the cluster center, where the galaxy density is high, they do not stand out in the galaxy number density maps. There are four X-ray groups where the brightest group galaxy has $\log \sigma > 2.4$. For the remaining four X-ray groups, the brightest galaxies are all quiescent ($\dn \geq 1.7$), but the $\sigma$s are not extreme ($2.2 < \log \sigma < 2.4$).

Interestingly, large $\sigma$ galaxies located outside $R_{200}$ (blue circles in Figure \ref{spatial}) often trace the potentially infalling substructures; these objects are excluded from the VDF measurements. For example, large $\sigma$ galaxies mark the center of each group in the cases of A2390, A1703, and RXJ 2129. Indeed, these large $\sigma$ galaxies are the brightest galaxies among the cluster members in the infalling substructures. 

However, some infalling groups do not contain these large $\sigma$ galaxies. A773, for example, has four potentially infalling substructures, but only two of these substructures have large $\sigma$ galaxies at their centers. The brightest galaxies in the other two substructures have $\sigma < 250~\kms$; these objects are not quiescent galaxies with $\dn < 1.5$. We note that the infalling groups may be at different evolutionary stages. For example, \citet{2019ApJ...871..129S} show that the galaxy composition of two infalling groups associated with the massive cluster A2029 differ significantly despite their similar mass. One group contains only a quiescent old population and the other consists mostly of star-forming objects. 

The association between large $\sigma$ galaxies and infalling structures suggests that many large $\sigma$ objects were central galaxies formed in a massive subhalo. Thus, we suggest that many large $\sigma$ galaxies within $R_{200}$ were originally central galaxies in accreted substructures. Previous work suggests a similar origin for massive galaxies in clusters based on the luminosity distribution of bright galaxies (e.g., \citealp{2004ApJ...617..879L, 2014ApJ...797...82L}). 

\subsection{Theoretical Guidance} \label{theory}

The large $\sigma$ excess in the cluster VDFs provides potentially important tests of massive galaxy cluster formation models. The variation in the abundance of massive galaxies that depends on cluster mass suggests a connection between the evolution of a cluster and its most massive members. Several studies identify trends in cluster luminosity functions that are similar to the trends in the velocity dispersion function \citep{2004ApJ...617..879L, 2012A&A...540A..90B}. These studies also suggest that massive galaxy formation may be tightly connected to the mass assembly history of the host clusters. 

\citet{2004ApJ...617..879L} propose that the main cluster BCGs grow through mergers with brightest group galaxies (BGGs) in accreted groups. They first show that luminous galaxies (brighter than $L_{*}$) are more abundant in less massive clusters. These luminous galaxies in less massive systems are bright and abundant enough to supply the luminosity needed for BCG growth as a more massive cluster accretes infalling groups. Mergers between the BCG at the cluster center and BGGs in infalling groups may be facilitated by the large size of BCGs and the shorter dynamical friction time scale that brings the BGGs into the cluster core. Based on N-body simulations, \citet{1998MNRAS.299..728T} show that large satellites fall into the cluster center and lose their identity thus producing preferential mergers between galaxies that formed as BCGs. 

\citet{2012A&A...540A..90B} suggest a similar scenario based on comparison between luminosity functions of relaxed and unrelaxed clusters. They also show that relaxed clusters contain fewer bright galaxies ($M_{r} < -22.5$), consistent with our observation of steeper VDF slopes for more massive clusters that are generally more relaxed (e.g., A2390). 

Extensive theoretical work focuses on BCG and massive galaxy formation. Mechanisms include cooling flows \citep{1977ApJ...215..723C}, galaxy mergers \citep{2005MNRAS.362..184B, 2006MNRAS.369.1081B, 2007MNRAS.375....2D, 2009ApJ...699L.178N, 2010ApJ...725.2312O, 2012MNRAS.424..747L, 2013ApJ...766...38L, 2016MNRAS.458.2371R, 2017MNRAS.467..661N}, and AGN feedback \citep{2010MNRAS.401.1099H, 2013MNRAS.436.1750R, 2017MNRAS.470.4186B}. Although massive galaxy formation is too complex to explain with a single mechanism, numerical simulations and semi-analytic models suggest that galaxy mergers account for most of the mass growth of early-type (quiescent) galaxies at $z < 2$. The `accreted' stellar mass fraction of galaxies is typically high ($60 - 80\%$, \citealp{2010ApJ...725.2312O}), and this fraction is higher for more massive galaxies \citep{2013ApJ...766...38L, 2016MNRAS.458.2371R}. 

``Dry" dissipationless mergers are the favored explanation for the mass growth of quiescent galaxies. Dry merger simulations match the observed properties of massive quiescent galaxies including sizes, shapes, stellar population ages, and kinematics \citep{2003ApJ...597L.117K, 2005MNRAS.362..184B, 2006MNRAS.369.1081B, 2006ApJ...636L..81N, 2007MNRAS.375....2D, 2013ApJ...766...38L, 2016MNRAS.456..300O, 2018ApJ...857...22L}. Dry merger simulations also account for BCG formation at $z < 1$ \citep{2007MNRAS.375....2D}, although some simulations suggest that the mass and age evolution of BCGs require the inclusion of some {\it in situ} star formation activity at low redshift (e.g., \citealp{2012ApJ...759...43T, 2018MNRAS.479.1125R}).

The mass assembly history of galaxy clusters directly affects the cluster VDFs. In the hierarchical structure formation model, galaxy clusters grow through accretion of surrounding substructures. Because these substructures tend to contain massive central galaxies (preprocessing, \citealp{1983ApJ...274..491B, 1998ApJ...496...39Z, 2007ApJ...671.1503M, 2014MNRAS.439.3564C, 2019MNRAS.485.2287B, 2019ApJ...885....6J}), the accretion history is clearly linked to the abundance of massive galaxies in clusters. 

Many numerical simulations show that the mass assembly history depends on the cluster halo mass. For a massive halo of $\sim 10^{15} M_{\odot}$, like the clusters in our sample, the mass of the halo has typically doubled since $z \sim 0.5$ \citep{2009ApJ...707..354Z, 2009MNRAS.398.1858M, 2010MNRAS.406.2267F, 2014MNRAS.445.1713V, 2016ApJ...818..188D, 2020arXiv200511562P}. However, more importantly, the mass accretion history is highly stochastic \citep{2016ApJ...818..188D, 2018MNRAS.477.4931H}; Figure 16 of \citet{2018MNRAS.477.4931H} demonstrates the stochastic mass accretion of massive cluster halos in the Millennium Simulations. Based on these simulations, we infer that a cluster acquires massive galaxies through the accretion of neighboring substructures although the frequency of this accretion varies from cluster to cluster. 

\citet{2014ApJ...786...79M} display an example of BCG and related massive galaxy evolution as a cluster grows by accreting surrounding structures. In their N-body simulation, a cluster forms at $z=0.54$ through a merger of three massive structures. The central galaxy in one of the progenitors migrates toward the center of cluster and becomes the BCG of the integrated cluster. At $z = 0$, the cluster is dynamically relaxed and harbours the descendant of this BCG at its center. Many massive galaxies ($> 10^{10} M_{\odot}$) also appear in the cluster center accompanying the clumpy dark matter substructures. 

Numerical simulations inform our understanding of the shapes of cluster VDFs because they allow us to trace the formation and evolution of massive (large $\sigma$) galaxies (e.g., \citealp{2014ApJ...786...79M}). Many studies investigate the luminosity increase and stellar mass growth of BCGs in massive cluster halos. However, estimation of galaxy luminosity and stellar mass in simulations is difficult due to the complexity of the underlying baryonic physics (e.g., feedback, \citealp{2013MNRAS.436.1750R, 2017MNRAS.470.4186B}). In contrast, estimation of the velocity dispersion of stellar particles in the simulations is less dependent on the details of the baryonic physics. For example, the impact of minor mergers on the central velocity dispersion is small compared to the effect on the stellar mass and/or luminosity \citep{2006ApJ...650..791C}. Direct estimation of $\sigma$ in the simulations and the VDF mimicking the observations \citep{2018ApJ...859...96Z} could enhance the understanding of massive galaxy evolution in clusters. Recent much larger simulations include enough massive clusters to enable this type of statistical study (e.g., Illustris TNG-300, \citealp{2018MNRAS.475..676S}). Numerical simulations do have some limitations including `subhalo overmerging' \citep{2017MNRAS.468..885V, 2018MNRAS.474.3043V}. Nonetheless they are an important guide because the velocity stellar velocity dispersion is a good proxy for the halo velocity dispersion.

\subsection{Observational Limitations and the Interpretation of Cluster VDFs} \label{limit}

A comprehensive interpretation of the observed cluster VDFs is challenging. A broad set of observations is required to develop a complete picture given the complex theoretical landscape. Issues range from the selection of the cluster sample to the inclusion of galaxy radii, galaxy shapes, the intracluster light, the distribution of massive cluster members, the role of AGN, and possibly spatially resolved spectroscopy of the cluster members. We briefly comment on several of these issues.

First, we note that our sample clusters were selected because they possess prominent strong lensing arcs in HST images. The strong lensing features require high projected mass concentration. The strong mass concentration toward the cluster center may be related to the cluster evolutionary stage. Thus, sampling based on lensing features could introduce a selection bias. Measuring VDFs of a much larger sample of clusters (e.g., HeCS-omnibus) would resolve this issue. The detection of similar behavior in luminosity functions (e.g., \citealp{2004ApJ...617..879L, 2012A&A...540A..90B}) provides some reassurance that the results we obtain are general.

We also lack size and shape measurements for the large $\sigma$ cluster members. The size and shape measurements can provide critical observational evidence of massive galaxy growth through dry mergers (e.g., \citealp{2007AJ....133.1741B}). Furthermore, robust measurements of galaxy sizes allows constraints on the mechanisms for quiescent galaxy evolution in massive clusters based on the fundamental plane (e.g., \citealp{2010MNRAS.407.2207F}) and/or the size-mass relation (e.g., \citealp{2017A&A...604A..54K}). In an ongoing investigation (Damjanov 2020, private communication) we have found that there are systematic differences between HST and Subaru HSC size measurements for cluster members that may depend on both the analysis technique and on the parameters chosen in a particular approach. We thus defer the discussion of sizes to a future investigation.

Recent theoretical and observation works suggest that intracluster light (ICL) is critical for elucidating the mass assembly of BCGs and their host clusters. Intracluster baryonic matter is distributed throughout clusters during active mass accretion. Because the BCG generally sits at or near the bottom of the cluster potential well, the ICL often appears as a halo around the BCG. Separating the BCG itself from the intracluster baryons/light is a challenge from both the theoretical and observational perspectives (e.g., \citealp{2003ApJ...594..172N, 2004ApJ...614L..33A, 2004ApJ...607L..83M, 2005ApJ...631L..41M}). Nonetheless, the amount of ICL and its distribution are a potential clue to understand the evolutionary track of the cluster. Indeed, \citet{2010MNRAS.405.1544D} show that ICL is prominent in clusters with an early formation time.

Despite the current observational limitations, the cluster VDF is an important tool. Measuring $\sigma$ has some clear advantages over measuring luminosity or stellar mass, particularly in the cluster environment. For example, deriving good photometry, and the related stellar mass based on the photometry in a crowded region is not straightforward \citep{2013MNRAS.436..697B}. In contrast, measurement of the central stellar velocity dispersion is relatively straightforward and is insensitive to crowding and to the choice of underlying models ({\it i.e.,} the stellar population models and star formation histories used for stellar mass estimation). 

Based on the VDFs and the spatial distribution of the massive cluster members, we sketch an evolutionary picture. The association between large $\sigma$ galaxies and infalling structures suggests that massive galaxies initially formed in massive group/cluster halos. As the main cluster incorporates other group/cluster halos, massive galaxies are accreted that were once central galaxies in their own halo. During this process, dry dissipationless mergers between the main BCG and other galaxies, including those accreted as members of infalling groups, occur. Dry dissipationless mergers account for a small increment in the BCG $\sigma$ \citep{2006ApJ...650..791C}. Both the number of large $\sigma$ galaxies and the $\sigma$ of the BCGs thus depend on the accretion history of the entire cluster.

This simple picture explains key features of the VDFs. At high velocity dispersion, cluster VDFs have a significant excess compared with the field VDF because large $\sigma$ galaxies form preferentially in clusters and neighboring substructures that are subsequently accreted. The cluster VDF excess is preserved or even enhanced when a cluster merges or accretes substructures containing large $\sigma$ galaxies. If a cluster predominantly accreted galaxies from the surrounding field environment, the cluster VDF excess would be diluted as the cluster evolved. In fact, the demonstration of `preprocessing' in less massive groups supports the idea that mass growth occurs in these environments. 

%=============================================================
\section{CONCLUSION} \label{conclusion}

The central stellar velocity dispersion of a galaxy is a remarkably good proxy for the central velocity dispersion of the host dark matter halo \citep{2013A&A...549A...7V, 2018ApJ...859...96Z, 2020arXiv200507122U}. Thus the velocity dispersion function can provide interesting insights about galaxy formation and evolution that complement the more traditional luminosity and stellar mass functions.

We derive velocity dispersion function for quiescent galaxies in 9 massive clusters in the redshift range $0.18 < z < 0.29$. All of these clusters produce strong lensing arcs in HST images. We use the same approach to measure the cluster VDFs as in previous field and cluster VDFs \citep{2017ApJS..229...20S, 2017ApJ...845...73S}. Compared to the field VDF, the cluster VDFs show a significant excess for galaxies with $\log \sigma > 2.4$ ($\sigma \gtrsim 250~\kms$). The VDFs of two local massive clusters, Coma and A2029, show a similar excess of large $\sigma$ galaxies. The large $\sigma$ excess generally includes the brightest cluster galaxy and a number of other large $\sigma$ galaxies. These objects are rare; they are bright, old (large $\dn$) galaxies. 

We draw these conclusions from dense spectroscopic surveys with MMT/Hectospec and SDSS. In each cluster, we identify $\sim 380$ ($\sim 183$) spectroscopic members (within $R_{200}$). We measure the spectroscopic properties of cluster members including the redshift, absolute magnitude, $\dn$, and the velocity dispersion within 3 kpc. We include these properties of the 3419 cluster members here. 

The spatial distribution of large $\sigma$ cluster members illustrates another interesting connection between massive galaxies and cluster substructure. Large $\sigma$ galaxies within $R_{200}$ are concentrated toward the cluster center. Interestingly, large $\sigma$ galaxies outside $R_{200}$, not included in the VDF measurements, trace the local density peaks of cluster members at $R > R_{200}$. These outlying density peaks are presumably infalling groups because they lie between the caustics that define the cluster population. The spatial distribution suggests that large $\sigma$ galaxies form in the centers of neighboring massive halos; they are eventually accreted onto the main cluster.

The slope of the cluster VDFs is a function of the cluster mass. The VDF of a more massive cluster tends to have a steeper slope for $\log \sigma > 2.2$. In other words, more massive clusters tend to have a smaller fraction of galaxies with large $\sigma$ than less massive clusters. \citet{2004ApJ...617..879L} and \citet{2012A&A...540A..90B} report a similar trend in the luminosity function of cluster galaxies.

The excess of large $\sigma$ galaxies in the cluster VDFs is a potentially important test of formation models for massive cluster members. The dependence of the VDF slope on cluster mass suggests a relationship between the evolution of a cluster and its massive members. However, a comprehensive interpretation of the observed cluster VDF is challenging because a larger set of observations including, for example, the sizes and shapes of cluster galaxies, are critical for a full comparison with galaxy evolutionary scenarios. 

We draw an evolutionary picture of galaxy clusters and their massive members based on the VDFs and theoretical models. A cluster grows by accreting neighboring massive halos containing their own massive central galaxies. The newly accreted halos fall toward the cluster center. During this process, the BCG grows through dry mergers with other massive galaxies in the accreted halos. Minor mergers are an additional source of BCG growth. This picture is consistent with many theoretical models. 

Future large spectroscopic surveys (e.g., the DESI survey, 4MOST, Subaru Prime Focus Spectrograph Survey) will allow us to extend the current VDFs to smaller velocity dispersion. They will also enable a direct evolutionary probe by enabling the determination of VDFs for clusters at different redshifts with a wide range of masses. The derivation of VDFs from large simulations would provide an important benchmark for these future observations. 

%=============================================================
\acknowledgments
We thank Perry Berlind and Michael Calkins for operating Hectospec. 
We also thank Susan Tokarz and Sean Moran for helping the data reduction. 
The Smithsonian Institution supported research of M.J.G.
This paper uses data products produced by the OIR Telescope Data Center, 
 supported by the Smithsonian Astrophysical Observatory.
J.S. gratefully acknowledges the support of the CfA Fellowship. 
The Smithsonian Institution supported the research of M.J.G. and D.F. 
AD acknowledges partial support from the Italian Ministry of Education, University and Research (MIUR) under the {\it Departments of Excellence} grant L.232/2016, and from the INFN grant InDark.
This research has made use of NASA's Astrophysics Data System Bibliographic Services. 
Funding for SDSS-III has been provided by the Alfred P. Sloan Foundation, 
 the Participating Institutions, the National Science Foundation, 
 and the U.S. Department of Energy Office of Science. 
The SDSS-III web site is http://www.sdss3.org/. 
SDSS-III is managed by the Astrophysical Research Consortium for 
 the Participating Institutions of the SDSS-III Collaboration 
 including the University of Arizona, 
 the Brazilian Participation Group, Brookhaven National Laboratory, 
 University of Cambridge, Carnegie Mellon University, 
 University of Florida, the French Participation Group, 
 the German Participation Group, Harvard University, 
 the Instituto de Astrofisica de Canarias, 
 the Michigan State/Notre Dame/ JINA Participation Group, 
 Johns Hopkins University, Lawrence Berkeley National Laboratory, 
 Max Planck Institute for Astrophysics, 
 Max Planck Institute for Extraterrestrial Physics, 
 New Mexico State University, New York University, Ohio State University, 
 Pennsylvania State University, University of Portsmouth, Princeton University, 
 the Spanish Participation Group, University of Tokyo, University of Utah, 
 Vanderbilt University, University of Virginia, University of Washington, and Yale University.
 
%\facilities{MMT:Hectospec} 

\bibliography{ms}{}

\begin{thebibliography}{}
\expandafter\ifx\csname natexlab\endcsname\relax\def\natexlab#1{#1}\fi
\providecommand{\url}[1]{\href{#1}{#1}}
\providecommand{\dodoi}[1]{doi:~\href{http://doi.org/#1}{\nolinkurl{#1}}}
\providecommand{\doeprint}[1]{\href{http://ascl.net/#1}{\nolinkurl{http://ascl.net/#1}}}
\providecommand{\doarXiv}[1]{\href{https://arxiv.org/abs/#1}{\nolinkurl{https://arxiv.org/abs/#1}}}

\bibitem[{{Abolfathi} {et~al.}(2018){Abolfathi}, {Aguado}, {Aguilar}, {Allende
  Prieto}, {Almeida}, {Ananna}, {Anders}, {Anderson}, {Andrews}, {Anguiano},
  {Arag{\'o}n-Salamanca}, {Argudo-Fern{\'a}ndez}, {Armengaud}, {Ata},
  {Aubourg}, {Avila-Reese}, {Badenes}, {Bailey}, {Balland}, {Barger},
  {Barrera-Ballesteros}, {Bartosz}, {Bastien}, {Bates}, {Baumgarten},
  {Bautista}, {Beaton}, {Beers}, {Belfiore}, {Bender}, {Bernardi}, {Bershady},
  {Beutler}, {Bird}, {Bizyaev}, {Blanc}, {Blanton}, {Blomqvist}, {Bolton},
  {Boquien}, {Borissova}, {Bovy}, {Bradna Diaz}, {Brandt}, {Brinkmann},
  {Brownstein}, {Bundy}, {Burgasser}, {Burtin}, {Busca}, {Ca{\~n}as},
  {Cano-D{\'\i}az}, {Cappellari}, {Carrera}, {Casey}, {Cervantes Sodi}, {Chen},
  {Cherinka}, {Chiappini}, {Choi}, {Chojnowski}, {Chuang}, {Chung}, {Clerc},
  {Cohen}, {Comerford}, {Comparat}, {Correa do Nascimento}, {da Costa},
  {Cousinou}, {Covey}, {Crane}, {Cruz-Gonzalez}, {Cunha}, {da Silva Ilha},
  {Damke}, {Darling}, {Davidson}, {Dawson}, {de Icaza Lizaola}, {de la
  Macorra}, {de la Torre}, {De Lee}, {de Sainte Agathe}, {Deconto Machado},
  {Dell'Agli}, {Delubac}, {Diamond-Stanic}, {Donor}, {Downes}, {Drory}, {du Mas
  des Bourboux}, {Duckworth}, {Dwelly}, {Dyer}, {Ebelke}, {Davis Eigenbrot},
  {Eisenstein}, {Elsworth}, {Emsellem}, {Eracleous}, {Erfanianfar},
  {Escoffier}, {Fan}, {Fern{\'a}ndez Alvar}, {Fernandez-Trincado}, {Fernand o
  Cirolini}, {Feuillet}, {Finoguenov}, {Fleming}, {Font-Ribera}, {Freischlad},
  {Frinchaboy}, {Fu}, {G{\'o}mez Maqueo Chew}, {Galbany}, {Garc{\'\i}a
  P{\'e}rez}, {Garcia-Dias}, {Garc{\'\i}a-Hern{\'a}ndez}, {Garma Oehmichen},
  {Gaulme}, {Gelfand }, {Gil-Mar{\'\i}n}, {Gillespie}, {Goddard}, {Gonz{\'a}lez
  Hern{\'a}ndez}, {Gonzalez-Perez}, {Grabowski}, {Green}, {Grier}, {Gueguen},
  {Guo}, {Guy}, {Hagen}, {Hall}, {Harding}, {Hasselquist}, {Hawley}, {Hayes},
  {Hearty}, {Hekker}, {Hernand ez}, {Hernandez Toledo}, {Hogg},
  {Holley-Bockelmann}, {Holtzman}, {Hou}, {Hsieh}, {Hunt}, {Hutchinson},
  {Hwang}, {Jimenez Angel}, {Johnson}, {Jones}, {J{\"o}nsson}, {Jullo}, {Khan},
  {Kinemuchi}, {Kirkby}, {Kirkpatrick}, {Kitaura}, {Knapp}, {Kneib},
  {Kollmeier}, {Lacerna}, {Lane}, {Lang}, {Law}, {Le Goff}, {Lee}, {Li}, {Li},
  {Lian}, {Liang}, {Lima}, {Lin}, {Long}, {Lucatello}, {Lundgren}, {Mackereth},
  {MacLeod}, {Mahadevan}, {Maia}, {Majewski}, {Manchado}, {Maraston},
  {Mariappan}, {Marques-Chaves}, {Masseron}, {Masters}, {McDermid}, {McGreer},
  {Melendez}, {Meneses-Goytia}, {Merloni}, {Merrifield}, {Meszaros}, {Meza},
  {Minchev}, {Minniti}, {Mueller}, {Muller-Sanchez}, {Muna}, {Mu{\~n}oz},
  {Myers}, {Nair}, {Nand ra}, {Ness}, {Newman}, {Nichol}, {Nidever},
  {Nitschelm}, {Noterdaeme}, {O'Connell}, {Oelkers}, {Oravetz}, {Oravetz},
  {Ort{\'\i}z}, {Osorio}, {Pace}, {Padilla}, {Palanque-Delabrouille},
  {Palicio}, {Pan}, {Pan}, {Parikh}, {P{\^a}ris}, {Park}, {Peirani},
  {Pellejero-Ibanez}, {Penny}, {Percival}, {Perez-Fournon}, {Petitjean},
  {Pieri}, {Pinsonneault}, {Pisani}, {Prada}, {Prakash}, {Queiroz}, {Raddick},
  {Raichoor}, {Barboza Rembold}, {Richstein}, {Riffel}, {Riffel}, {Rix},
  {Robin}, {Rodr{\'\i}guez Torres}, {Rom{\'a}n-Z{\'u}{\~n}iga}, {Ross},
  {Rossi}, {Ruan}, {Ruggeri}, {Ruiz}, {Salvato}, {S{\'a}nchez}, {S{\'a}nchez},
  {Sanchez Almeida}, {S{\'a}nchez-Gallego}, {Santana Rojas}, {Santiago},
  {Schiavon}, {Schimoia}, {Schlafly}, {Schlegel}, {Schneider}, {Schuster},
  {Schwope}, {Seo}, {Serenelli}, {Shen}, {Shen}, {Shetrone}, {Shull}, {Silva
  Aguirre}, {Simon}, {Skrutskie}, {Slosar}, {Smethurst}, {Smith}, {Sobeck},
  {Somers}, {Souter}, {Souto}, {Spindler}, {Stark}, {Stassun}, {Steinmetz},
  {Stello}, {Storchi-Bergmann}, {Streblyanska}, {Stringfellow}, {Su{\'a}rez},
  {Sun}, {Szigeti}, {Taghizadeh-Popp}, {Talbot}, {Tang}, {Tao}, {Tayar},
  {Tembe}, {Teske}, {Thakar}, {Thomas}, {Tissera}, {Tojeiro}, {Tremonti},
  {Troup}, {Urry}, {Valenzuela}, {van den Bosch}, {Vargas-Gonz{\'a}lez},
  {Vargas-Maga{\~n}a}, {Vazquez}, {Villanova}, {Vogt}, {Wake}, {Wang},
  {Weaver}, {Weijmans}, {Weinberg}, {Westfall}, {Whelan}, {Wilcots}, {Wild},
  {Williams}, {Wilson}, {Wood-Vasey}, {Wylezalek}, {Xiao}, {Yan}, {Yang},
  {Ybarra}, {Y{\`e}che}, {Zakamska}, {Zamora}, {Zarrouk}, {Zasowski}, {Zhang},
  {Zhao}, {Zhao}, {Zheng}, {Zheng}, {Zhou}, {Zhu}, {Zinn}, \&
  {Zou}}]{2018ApJS..235...42A}
{Abolfathi}, B., {Aguado}, D.~S., {Aguilar}, G., {et~al.} 2018, \apjs, 235, 42,
  \dodoi{10.3847/1538-4365/aa9e8a}

\bibitem[{{Arnaboldi} {et~al.}(2004){Arnaboldi}, {Gerhard}, {Aguerri},
  {Freeman}, {Napolitano}, {Okamura}, \& {Yasuda}}]{2004ApJ...614L..33A}
{Arnaboldi}, M., {Gerhard}, O., {Aguerri}, J. A.~L., {et~al.} 2004, \apjl, 614,
  L33, \dodoi{10.1086/425417}

\bibitem[{{Bah{\'e}} {et~al.}(2017){Bah{\'e}}, {Barnes}, {Dalla Vecchia},
  {Kay}, {White}, {McCarthy}, {Schaye}, {Bower}, {Crain}, {Theuns}, {Jenkins},
  {McGee}, {Schaller}, {Thomas}, \& {Trayford}}]{2017MNRAS.470.4186B}
{Bah{\'e}}, Y.~M., {Barnes}, D.~J., {Dalla Vecchia}, C., {et~al.} 2017, \mnras,
  470, 4186, \dodoi{10.1093/mnras/stx1403}

\bibitem[{{Bah{\'e}} {et~al.}(2019){Bah{\'e}}, {Schaye}, {Barnes}, {Dalla
  Vecchia}, {Kay}, {Bower}, {Hoekstra}, {McGee}, \&
  {Theuns}}]{2019MNRAS.485.2287B}
{Bah{\'e}}, Y.~M., {Schaye}, J., {Barnes}, D.~J., {et~al.} 2019, \mnras, 485,
  2287, \dodoi{10.1093/mnras/stz361}

\bibitem[{{Balogh} {et~al.}(1999){Balogh}, {Morris}, {Yee}, {Carlberg}, \&
  {Ellingson}}]{1999ApJ...527...54B}
{Balogh}, M.~L., {Morris}, S.~L., {Yee}, H.~K.~C., {Carlberg}, R.~G., \&
  {Ellingson}, E. 1999, \apj, 527, 54, \dodoi{10.1086/308056}

\bibitem[{{Barrena} {et~al.}(2012){Barrena}, {Girardi}, {Boschin}, \&
  {Mardirossian}}]{2012A&A...540A..90B}
{Barrena}, R., {Girardi}, M., {Boschin}, W., \& {Mardirossian}, F. 2012, \aap,
  540, A90, \dodoi{10.1051/0004-6361/201118586}

\bibitem[{{Beers} \& {Geller}(1983)}]{1983ApJ...274..491B}
{Beers}, T.~C., \& {Geller}, M.~J. 1983, \apj, 274, 491, \dodoi{10.1086/161463}

\bibitem[{{Behroozi} {et~al.}(2019){Behroozi}, {Wechsler}, {Hearin}, \&
  {Conroy}}]{2019MNRAS.488.3143B}
{Behroozi}, P., {Wechsler}, R.~H., {Hearin}, A.~P., \& {Conroy}, C. 2019,
  \mnras, 488, 3143, \dodoi{10.1093/mnras/stz1182}

\bibitem[{{Bernardi} {et~al.}(2007){Bernardi}, {Hyde}, {Sheth}, {Miller}, \&
  {Nichol}}]{2007AJ....133.1741B}
{Bernardi}, M., {Hyde}, J.~B., {Sheth}, R.~K., {Miller}, C.~J., \& {Nichol},
  R.~C. 2007, \aj, 133, 1741, \dodoi{10.1086/511783}

\bibitem[{{Bernardi} {et~al.}(2013){Bernardi}, {Meert}, {Sheth}, {Vikram},
  {Huertas-Company}, {Mei}, \& {Shankar}}]{2013MNRAS.436..697B}
{Bernardi}, M., {Meert}, A., {Sheth}, R.~K., {et~al.} 2013, \mnras, 436, 697,
  \dodoi{10.1093/mnras/stt1607}

\bibitem[{{Blanton} \& {Moustakas}(2009)}]{2009ARA&A..47..159B}
{Blanton}, M.~R., \& {Moustakas}, J. 2009, \araa, 47, 159,
  \dodoi{10.1146/annurev-astro-082708-101734}

\bibitem[{{Bogd{\'a}n} \& {Goulding}(2015)}]{2015ApJ...800..124B}
{Bogd{\'a}n}, {\'A}., \& {Goulding}, A.~D. 2015, \apj, 800, 124,
  \dodoi{10.1088/0004-637X/800/2/124}

\bibitem[{{Boselli} \& {Gavazzi}(2006)}]{2006PASP..118..517B}
{Boselli}, A., \& {Gavazzi}, G. 2006, \pasp, 118, 517, \dodoi{10.1086/500691}

\bibitem[{{Boylan-Kolchin} {et~al.}(2005){Boylan-Kolchin}, {Ma}, \&
  {Quataert}}]{2005MNRAS.362..184B}
{Boylan-Kolchin}, M., {Ma}, C.-P., \& {Quataert}, E. 2005, \mnras, 362, 184,
  \dodoi{10.1111/j.1365-2966.2005.09278.x}

\bibitem[{{Boylan-Kolchin} {et~al.}(2006){Boylan-Kolchin}, {Ma}, \&
  {Quataert}}]{2006MNRAS.369.1081B}
---. 2006, \mnras, 369, 1081, \dodoi{10.1111/j.1365-2966.2006.10379.x}

\bibitem[{{Calvi} {et~al.}(2013){Calvi}, {Poggianti}, {Vulcani}, \&
  {Fasano}}]{2013MNRAS.432.3141C}
{Calvi}, R., {Poggianti}, B.~M., {Vulcani}, B., \& {Fasano}, G. 2013, \mnras,
  432, 3141, \dodoi{10.1093/mnras/stt667}

\bibitem[{{Cappellari} \& {Emsellem}(2004)}]{2004PASP..116..138C}
{Cappellari}, M., \& {Emsellem}, E. 2004, \pasp, 116, 138,
  \dodoi{10.1086/381875}

\bibitem[{{Cappellari} {et~al.}(2006){Cappellari}, {Bacon}, {Bureau}, {Damen},
  {Davies}, {de Zeeuw}, {Emsellem}, {Falc{\'o}n-Barroso}, {Krajnovi{\'c}},
  {Kuntschner}, {McDermid}, {Peletier}, {Sarzi}, {van den Bosch}, \& {van de
  Ven}}]{2006MNRAS.366.1126C}
{Cappellari}, M., {Bacon}, R., {Bureau}, M., {et~al.} 2006, \mnras, 366, 1126,
  \dodoi{10.1111/j.1365-2966.2005.09981.x}

\bibitem[{{Chae}(2010)}]{2010MNRAS.402.2031C}
{Chae}, K.-H. 2010, \mnras, 402, 2031, \dodoi{10.1111/j.1365-2966.2009.16073.x}

\bibitem[{{Choi} {et~al.}(2007){Choi}, {Park}, \&
  {Vogeley}}]{2007ApJ...658..884C}
{Choi}, Y.-Y., {Park}, C., \& {Vogeley}, M.~S. 2007, \apj, 658, 884,
  \dodoi{10.1086/511060}

\bibitem[{{Christlein} \& {Zabludoff}(2003)}]{2003ApJ...591..764C}
{Christlein}, D., \& {Zabludoff}, A.~I. 2003, \apj, 591, 764,
  \dodoi{10.1086/375529}

\bibitem[{{Cowie} \& {Binney}(1977)}]{1977ApJ...215..723C}
{Cowie}, L.~L., \& {Binney}, J. 1977, \apj, 215, 723, \dodoi{10.1086/155406}

\bibitem[{{Cox} {et~al.}(2006){Cox}, {Dutta}, {Di Matteo}, {Hernquist},
  {Hopkins}, {Robertson}, \& {Springel}}]{2006ApJ...650..791C}
{Cox}, T.~J., {Dutta}, S.~N., {Di Matteo}, T., {et~al.} 2006, \apj, 650, 791,
  \dodoi{10.1086/507474}

\bibitem[{{Cybulski} {et~al.}(2014){Cybulski}, {Yun}, {Fazio}, \&
  {Gutermuth}}]{2014MNRAS.439.3564C}
{Cybulski}, R., {Yun}, M.~S., {Fazio}, G.~G., \& {Gutermuth}, R.~A. 2014,
  \mnras, 439, 3564, \dodoi{10.1093/mnras/stu200}

\bibitem[{{Davidzon} {et~al.}(2016){Davidzon}, {Cucciati}, {Bolzonella}, {De
  Lucia}, {Zamorani}, {Arnouts}, {Moutard}, {Ilbert}, {Garilli}, {Scodeggio},
  {Guzzo}, {Abbas}, {Adami}, {Bel}, {Bottini}, {Branchini}, {Cappi}, {Coupon},
  {de la Torre}, {Di Porto}, {Fritz}, {Franzetti}, {Fumana}, {Granett},
  {Guennou}, {Iovino}, {Krywult}, {Le Brun}, {Le F{\`e}vre}, {Maccagni},
  {Ma{\l}ek}, {Marulli}, {McCracken}, {Mellier}, {Moscardini}, {Polletta},
  {Pollo}, {Tasca}, {Tojeiro}, {Vergani}, \&
  {Zanichelli}}]{2016A&A...586A..23D}
{Davidzon}, I., {Cucciati}, O., {Bolzonella}, M., {et~al.} 2016, \aap, 586,
  A23, \dodoi{10.1051/0004-6361/201527129}

\bibitem[{{Dawson} {et~al.}(2013){Dawson}, {Schlegel}, {Ahn}, {Anderson},
  {Aubourg}, {Bailey}, {Barkhouser}, {Bautista}, {Beifiori}, {Berlind},
  {Bhardwaj}, {Bizyaev}, {Blake}, {Blanton}, {Blomqvist}, {Bolton}, {Borde},
  {Bovy}, {Brandt}, {Brewington}, {Brinkmann}, {Brown}, {Brownstein}, {Bundy},
  {Busca}, {Carithers}, {Carnero}, {Carr}, {Chen}, {Comparat}, {Connolly},
  {Cope}, {Croft}, {Cuesta}, {da Costa}, {Davenport}, {Delubac}, {de Putter},
  {Dhital}, {Ealet}, {Ebelke}, {Eisenstein}, {Escoffier}, {Fan}, {Filiz Ak},
  {Finley}, {Font-Ribera}, {G{\'e}nova-Santos}, {Gunn}, {Guo}, {Haggard},
  {Hall}, {Hamilton}, {Harris}, {Harris}, {Ho}, {Hogg}, {Holder}, {Honscheid},
  {Huehnerhoff}, {Jordan}, {Jordan}, {Kauffmann}, {Kazin}, {Kirkby}, {Klaene},
  {Kneib}, {Le Goff}, {Lee}, {Long}, {Loomis}, {Lundgren}, {Lupton}, {Maia},
  {Makler}, {Malanushenko}, {Malanushenko}, {Mandelbaum}, {Manera}, {Maraston},
  {Margala}, {Masters}, {McBride}, {McDonald}, {McGreer}, {McMahon}, {Mena},
  {Miralda-Escud{\'e}}, {Montero-Dorta}, {Montesano}, {Muna}, {Myers},
  {Naugle}, {Nichol}, {Noterdaeme}, {Nuza}, {Olmstead}, {Oravetz}, {Oravetz},
  {Owen}, {Padmanabhan}, {Palanque-Delabrouille}, {Pan}, {Parejko},
  {P{\^a}ris}, {Percival}, {P{\'e}rez-Fournon}, {P{\'e}rez-R{\`a}fols},
  {Petitjean}, {Pfaffenberger}, {Pforr}, {Pieri}, {Prada}, {Price-Whelan},
  {Raddick}, {Rebolo}, {Rich}, {Richards}, {Rockosi}, {Roe}, {Ross}, {Ross},
  {Rossi}, {Rubi{\~n}o-Martin}, {Samushia}, {S{\'a}nchez}, {Sayres}, {Schmidt},
  {Schneider}, {Sc{\'o}ccola}, {Seo}, {Shelden}, {Sheldon}, {Shen}, {Shu},
  {Slosar}, {Smee}, {Snedden}, {Stauffer}, {Steele}, {Strauss}, {Streblyanska},
  {Suzuki}, {Swanson}, {Tal}, {Tanaka}, {Thomas}, {Tinker}, {Tojeiro},
  {Tremonti}, {Vargas Maga{\~n}a}, {Verde}, {Viel}, {Wake}, {Watson}, {Weaver},
  {Weinberg}, {Weiner}, {West}, {White}, {Wood-Vasey}, {Yeche}, {Zehavi},
  {Zhao}, \& {Zheng}}]{2013AJ....145...10D}
{Dawson}, K.~S., {Schlegel}, D.~J., {Ahn}, C.~P., {et~al.} 2013, \aj, 145, 10,
  \dodoi{10.1088/0004-6256/145/1/10}

\bibitem[{{De Boni} {et~al.}(2016){De Boni}, {Serra}, {Diaferio}, {Giocoli}, \&
  {Baldi}}]{2016ApJ...818..188D}
{De Boni}, C., {Serra}, A.~L., {Diaferio}, A., {Giocoli}, C., \& {Baldi}, M.
  2016, \apj, 818, 188, \dodoi{10.3847/0004-637X/818/2/188}

\bibitem[{{De Lucia} \& {Blaizot}(2007)}]{2007MNRAS.375....2D}
{De Lucia}, G., \& {Blaizot}, J. 2007, \mnras, 375, 2,
  \dodoi{10.1111/j.1365-2966.2006.11287.x}

\bibitem[{{Dey} {et~al.}(2019){Dey}, {Schlegel}, {Lang}, {Blum}, {Burleigh},
  {Fan}, {Findlay}, {Finkbeiner}, {Herrera}, {Juneau}, {Landriau}, {Levi},
  {McGreer}, {Meisner}, {Myers}, {Moustakas}, {Nugent}, {Patej}, {Schlafly},
  {Walker}, {Valdes}, {Weaver}, {Y{\`e}che}, {Zou}, {Zhou}, {Abareshi},
  {Abbott}, {Abolfathi}, {Aguilera}, {Alam}, {Allen}, {Alvarez}, {Annis},
  {Ansarinejad}, {Aubert}, {Beechert}, {Bell}, {BenZvi}, {Beutler}, {Bielby},
  {Bolton}, {Brice{\~n}o}, {Buckley-Geer}, {Butler}, {Calamida}, {Carlberg},
  {Carter}, {Casas}, {Castander}, {Choi}, {Comparat}, {Cukanovaite}, {Delubac},
  {DeVries}, {Dey}, {Dhungana}, {Dickinson}, {Ding}, {Donaldson}, {Duan},
  {Duckworth}, {Eftekharzadeh}, {Eisenstein}, {Etourneau}, {Fagrelius},
  {Farihi}, {Fitzpatrick}, {Font-Ribera}, {Fulmer}, {G{\"a}nsicke},
  {Gaztanaga}, {George}, {Gerdes}, {Gontcho}, {Gorgoni}, {Green}, {Guy},
  {Harmer}, {Hernand ez}, {Honscheid}, {Huang}, {James}, {Jannuzi}, {Jiang},
  {Joyce}, {Karcher}, {Karkar}, {Kehoe}, {Kneib}, {Kueter-Young}, {Lan},
  {Lauer}, {Le Guillou}, {Le Van Suu}, {Lee}, {Lesser}, {Perreault Levasseur},
  {Li}, {Mann}, {Marshall}, {Mart{\'\i}nez-V{\'a}zquez}, {Martini}, {du Mas des
  Bourboux}, {McManus}, {Meier}, {M{\'e}nard}, {Metcalfe},
  {Mu{\~n}oz-Guti{\'e}rrez}, {Najita}, {Napier}, {Narayan}, {Newman}, {Nie},
  {Nord}, {Norman}, {Olsen}, {Paat}, {Palanque-Delabrouille}, {Peng},
  {Poppett}, {Poremba}, {Prakash}, {Rabinowitz}, {Raichoor}, {Rezaie},
  {Robertson}, {Roe}, {Ross}, {Ross}, {Rudnick}, {Safonova}, {Saha},
  {S{\'a}nchez}, {Savary}, {Schweiker}, {Scott}, {Seo}, {Shan}, {Silva},
  {Slepian}, {Soto}, {Sprayberry}, {Staten}, {Stillman}, {Stupak}, {Summers},
  {Sien Tie}, {Tirado}, {Vargas-Maga{\~n}a}, {Vivas}, {Wechsler}, {Williams},
  {Yang}, {Yang}, {Yapici}, {Zaritsky}, {Zenteno}, {Zhang}, {Zhang}, {Zhou}, \&
  {Zhou}}]{2019AJ....157..168D}
{Dey}, A., {Schlegel}, D.~J., {Lang}, D., {et~al.} 2019, \aj, 157, 168,
  \dodoi{10.3847/1538-3881/ab089d}

\bibitem[{{Diaferio}(1999)}]{1999MNRAS.309..610D}
{Diaferio}, A. 1999, \mnras, 309, 610, \dodoi{10.1046/j.1365-8711.1999.02864.x}

\bibitem[{{Diaferio} \& {Geller}(1997)}]{1997ApJ...481..633D}
{Diaferio}, A., \& {Geller}, M.~J. 1997, \apj, 481, 633, \dodoi{10.1086/304075}

\bibitem[{{Dolag} {et~al.}(2010){Dolag}, {Murante}, \&
  {Borgani}}]{2010MNRAS.405.1544D}
{Dolag}, K., {Murante}, G., \& {Borgani}, S. 2010, \mnras, 405, 1544,
  \dodoi{10.1111/j.1365-2966.2010.16583.x}

\bibitem[{{Dressler}(1980)}]{1980ApJ...236..351D}
{Dressler}, A. 1980, \apj, 236, 351, \dodoi{10.1086/157753}

\bibitem[{{Fabricant} {et~al.}(2013){Fabricant}, {Chilingarian}, {Hwang},
  {Kurtz}, {Geller}, {Del'Antonio}, \& {Rines}}]{2013PASP..125.1362F}
{Fabricant}, D., {Chilingarian}, I., {Hwang}, H.~S., {et~al.} 2013, \pasp, 125,
  1362, \dodoi{10.1086/673499}

\bibitem[{{Fabricant} {et~al.}(2005){Fabricant}, {Fata}, {Roll}, {Hertz},
  {Caldwell}, {Gauron}, {Geary}, {McLeod}, {Szentgyorgyi}, {Zajac}, {Kurtz},
  {Barberis}, {Bergner}, {Brown}, {Conroy}, {Eng}, {Geller}, {Goddard},
  {Honsa}, {Mueller}, {Mink}, {Ordway}, {Tokarz}, {Woods}, {Wyatt}, {Epps}, \&
  {Dell'Antonio}}]{2005PASP..117.1411F}
{Fabricant}, D., {Fata}, R., {Roll}, J., {et~al.} 2005, \pasp, 117, 1411,
  \dodoi{10.1086/497385}

\bibitem[{{Fabricant} {et~al.}(2008){Fabricant}, {Kurtz}, {Geller}, {Caldwell},
  {Woods}, \& {Dell'Antonio}}]{2008PASP..120.1222F}
{Fabricant}, D.~G., {Kurtz}, M.~J., {Geller}, M.~J., {et~al.} 2008, \pasp, 120,
  1222, \dodoi{10.1086/593023}

\bibitem[{{Fakhouri} {et~al.}(2010){Fakhouri}, {Ma}, \&
  {Boylan-Kolchin}}]{2010MNRAS.406.2267F}
{Fakhouri}, O., {Ma}, C.-P., \& {Boylan-Kolchin}, M. 2010, \mnras, 406, 2267,
  \dodoi{10.1111/j.1365-2966.2010.16859.x}

\bibitem[{{Fraix-Burnet} {et~al.}(2010){Fraix-Burnet}, {Dugu{\'e}},
  {Chattopadhyay}, {Chattopadhyay}, \& {Davoust}}]{2010MNRAS.407.2207F}
{Fraix-Burnet}, D., {Dugu{\'e}}, M., {Chattopadhyay}, T., {Chattopadhyay},
  A.~K., \& {Davoust}, E. 2010, \mnras, 407, 2207,
  \dodoi{10.1111/j.1365-2966.2010.17097.x}

\bibitem[{{Geller} {et~al.}(2014){Geller}, {Hwang}, {Diaferio}, {Kurtz}, {Coe},
  \& {Rines}}]{2014ApJ...783...52G}
{Geller}, M.~J., {Hwang}, H.~S., {Diaferio}, A., {et~al.} 2014, \apj, 783, 52,
  \dodoi{10.1088/0004-637X/783/1/52}

\bibitem[{{Gunn} {et~al.}(2006){Gunn}, {Siegmund}, {Mannery}, {Owen}, {Hull},
  {Leger}, {Carey}, {Knapp}, {York}, {Boroski}, {Kent}, {Lupton}, {Rockosi},
  {Evans}, {Waddell}, {Anderson}, {Annis}, {Barentine}, {Bartoszek}, {Bastian},
  {Bracker}, {Brewington}, {Briegel}, {Brinkmann}, {Brown}, {Carr},
  {Czarapata}, {Drennan}, {Dombeck}, {Federwitz}, {Gillespie}, {Gonzales},
  {Hansen}, {Harvanek}, {Hayes}, {Jordan}, {Kinney}, {Klaene}, {Kleinman},
  {Kron}, {Kresinski}, {Lee}, {Limmongkol}, {Lindenmeyer}, {Long}, {Loomis},
  {McGehee}, {Mantsch}, {Neilsen}, {Neswold}, {Newman}, {Nitta}, {Peoples},
  {Pier}, {Prieto}, {Prosapio}, {Rivetta}, {Schneider}, {Snedden}, \&
  {Wang}}]{2006AJ....131.2332G}
{Gunn}, J.~E., {Siegmund}, W.~A., {Mannery}, E.~J., {et~al.} 2006, \aj, 131,
  2332, \dodoi{10.1086/500975}

\bibitem[{{Haines} {et~al.}(2013){Haines}, {Pereira}, {Smith}, {Egami},
  {Sanderson}, {Babul}, {Finoguenov}, {Merluzzi}, {Busarello}, {Rawle}, \&
  {Okabe}}]{2013ApJ...775..126H}
{Haines}, C.~P., {Pereira}, M.~J., {Smith}, G.~P., {et~al.} 2013, \apj, 775,
  126, \dodoi{10.1088/0004-637X/775/2/126}

\bibitem[{{Haines} {et~al.}(2015){Haines}, {Pereira}, {Smith}, {Egami},
  {Babul}, {Finoguenov}, {Ziparo}, {McGee}, {Rawle}, {Okabe}, \&
  {Moran}}]{2015ApJ...806..101H}
---. 2015, \apj, 806, 101, \dodoi{10.1088/0004-637X/806/1/101}

\bibitem[{{Haines} {et~al.}(2018){Haines}, {Finoguenov}, {Smith}, {Babul},
  {Egami}, {Mazzotta}, {Okabe}, {Pereira}, {Bianconi}, {McGee}, {Ziparo},
  {Campusano}, \& {Loyola}}]{2018MNRAS.477.4931H}
{Haines}, C.~P., {Finoguenov}, A., {Smith}, G.~P., {et~al.} 2018, \mnras, 477,
  4931, \dodoi{10.1093/mnras/sty651}

\bibitem[{{Hasan} \& {Crocker}(2019)}]{2019arXiv190400486H}
{Hasan}, F., \& {Crocker}, A. 2019, arXiv e-prints, arXiv:1904.00486.
\newblock \doarXiv{1904.00486}

\bibitem[{{Hashimoto} {et~al.}(2007){Hashimoto}, {B{\"o}hringer}, {Henry},
  {Hasinger}, \& {Szokoly}}]{2007A&A...467..485H}
{Hashimoto}, Y., {B{\"o}hringer}, H., {Henry}, J.~P., {Hasinger}, G., \&
  {Szokoly}, G. 2007, \aap, 467, 485, \dodoi{10.1051/0004-6361:20065125}

\bibitem[{{Hashimoto} {et~al.}(2014){Hashimoto}, {Henry}, \&
  {Boehringer}}]{2014MNRAS.440..588H}
{Hashimoto}, Y., {Henry}, J.~P., \& {Boehringer}, H. 2014, \mnras, 440, 588,
  \dodoi{10.1093/mnras/stu311}

\bibitem[{{Hopkins} {et~al.}(2010){Hopkins}, {Bundy}, {Hernquist}, {Wuyts}, \&
  {Cox}}]{2010MNRAS.401.1099H}
{Hopkins}, P.~F., {Bundy}, K., {Hernquist}, L., {Wuyts}, S., \& {Cox}, T.~J.
  2010, \mnras, 401, 1099, \dodoi{10.1111/j.1365-2966.2009.15699.x}

\bibitem[{{Hwang} {et~al.}(2012){Hwang}, {Geller}, {Diaferio}, \&
  {Rines}}]{2012ApJ...752...64H}
{Hwang}, H.~S., {Geller}, M.~J., {Diaferio}, A., \& {Rines}, K.~J. 2012, \apj,
  752, 64, \dodoi{10.1088/0004-637X/752/1/64}

\bibitem[{{Hwang} {et~al.}(2014){Hwang}, {Geller}, {Diaferio}, {Rines}, \&
  {Zahid}}]{2014ApJ...797..106H}
{Hwang}, H.~S., {Geller}, M.~J., {Diaferio}, A., {Rines}, K.~J., \& {Zahid},
  H.~J. 2014, \apj, 797, 106, \dodoi{10.1088/0004-637X/797/2/106}

\bibitem[{{Just} {et~al.}(2019){Just}, {Kirby}, {Zaritsky}, {Rudnick},
  {Desjardins}, {Cool}, {Moustakas}, {Clowe}, {De Lucia},
  {Arag{\'o}n-Salamanca}, {Desai}, {Finn}, {Halliday}, {Jablonka}, {Mann},
  {Poggianti}, {Bian}, \& {Liebst}}]{2019ApJ...885....6J}
{Just}, D.~W., {Kirby}, M., {Zaritsky}, D., {et~al.} 2019, \apj, 885, 6,
  \dodoi{10.3847/1538-4357/ab44a0}

\bibitem[{{Kauffmann} {et~al.}(2004){Kauffmann}, {White}, {Heckman},
  {M{\'e}nard}, {Brinchmann}, {Charlot}, {Tremonti}, \&
  {Brinkmann}}]{2004MNRAS.353..713K}
{Kauffmann}, G., {White}, S. D.~M., {Heckman}, T.~M., {et~al.} 2004, \mnras,
  353, 713, \dodoi{10.1111/j.1365-2966.2004.08117.x}

\bibitem[{{Kauffmann} {et~al.}(2003){Kauffmann}, {Heckman}, {White}, {Charlot},
  {Tremonti}, {Brinchmann}, {Bruzual}, {Peng}, {Seibert}, {Bernardi},
  {Blanton}, {Brinkmann}, {Castander}, {Cs{\'a}bai}, {Fukugita}, {Ivezic},
  {Munn}, {Nichol}, {Padmanabhan}, {Thakar}, {Weinberg}, \&
  {York}}]{2003MNRAS.341...33K}
{Kauffmann}, G., {Heckman}, T.~M., {White}, S. D.~M., {et~al.} 2003, \mnras,
  341, 33, \dodoi{10.1046/j.1365-8711.2003.06291.x}

\bibitem[{{Khochfar} \& {Burkert}(2003)}]{2003ApJ...597L.117K}
{Khochfar}, S., \& {Burkert}, A. 2003, \apjl, 597, L117, \dodoi{10.1086/379845}

\bibitem[{{Klypin} {et~al.}(1999){Klypin}, {Gottl{\"o}ber}, {Kravtsov}, \&
  {Khokhlov}}]{1999ApJ...516..530K}
{Klypin}, A., {Gottl{\"o}ber}, S., {Kravtsov}, A.~V., \& {Khokhlov}, A.~M.
  1999, \apj, 516, 530, \dodoi{10.1086/307122}

\bibitem[{{Koleva} {et~al.}(2009){Koleva}, {Prugniel}, {Bouchard}, \&
  {Wu}}]{2009A&A...501.1269K}
{Koleva}, M., {Prugniel}, P., {Bouchard}, A., \& {Wu}, Y. 2009, \aap, 501,
  1269, \dodoi{10.1051/0004-6361/200811467}

\bibitem[{{Kuchner} {et~al.}(2017){Kuchner}, {Ziegler}, {Verdugo}, {Bamford},
  \& {H{\"a}u{\ss}ler}}]{2017A&A...604A..54K}
{Kuchner}, U., {Ziegler}, B., {Verdugo}, M., {Bamford}, S., \&
  {H{\"a}u{\ss}ler}, B. 2017, \aap, 604, A54,
  \dodoi{10.1051/0004-6361/201630252}

\bibitem[{{Lapi} {et~al.}(2018){Lapi}, {Pantoni}, {Zanisi}, {Shi}, {Mancuso},
  {Massardi}, {Shankar}, {Bressan}, \& {Danese}}]{2018ApJ...857...22L}
{Lapi}, A., {Pantoni}, L., {Zanisi}, L., {et~al.} 2018, \apj, 857, 22,
  \dodoi{10.3847/1538-4357/aab6af}

\bibitem[{{Laporte} {et~al.}(2012){Laporte}, {White}, {Naab}, {Ruszkowski}, \&
  {Springel}}]{2012MNRAS.424..747L}
{Laporte}, C. F.~P., {White}, S. D.~M., {Naab}, T., {Ruszkowski}, M., \&
  {Springel}, V. 2012, \mnras, 424, 747,
  \dodoi{10.1111/j.1365-2966.2012.21262.x}

\bibitem[{{Lauer} {et~al.}(2014){Lauer}, {Postman}, {Strauss}, {Graves}, \&
  {Chisari}}]{2014ApJ...797...82L}
{Lauer}, T.~R., {Postman}, M., {Strauss}, M.~A., {Graves}, G.~J., \& {Chisari},
  N.~E. 2014, \apj, 797, 82, \dodoi{10.1088/0004-637X/797/2/82}

\bibitem[{{Le Borgne} {et~al.}(2004){Le Borgne}, {Rocca-Volmerange},
  {Prugniel}, {Lan{\c{c}}on}, {Fioc}, \& {Soubiran}}]{2004A&A...425..881L}
{Le Borgne}, D., {Rocca-Volmerange}, B., {Prugniel}, P., {et~al.} 2004, \aap,
  425, 881, \dodoi{10.1051/0004-6361:200400044}

\bibitem[{{Lee} \& {Yi}(2013)}]{2013ApJ...766...38L}
{Lee}, J., \& {Yi}, S.~K. 2013, \apj, 766, 38,
  \dodoi{10.1088/0004-637X/766/1/38}

\bibitem[{{Lin} \& {Mohr}(2004)}]{2004ApJ...617..879L}
{Lin}, Y.-T., \& {Mohr}, J.~J. 2004, \apj, 617, 879, \dodoi{10.1086/425412}

\bibitem[{{Maraston} \& {Str{\"o}mb{\"a}ck}(2011)}]{2011MNRAS.418.2785M}
{Maraston}, C., \& {Str{\"o}mb{\"a}ck}, G. 2011, \mnras, 418, 2785,
  \dodoi{10.1111/j.1365-2966.2011.19738.x}

\bibitem[{{Martel} {et~al.}(2014){Martel}, {Robichaud}, \&
  {Barai}}]{2014ApJ...786...79M}
{Martel}, H., {Robichaud}, F., \& {Barai}, P. 2014, \apj, 786, 79,
  \dodoi{10.1088/0004-637X/786/2/79}

\bibitem[{{McBride} {et~al.}(2009){McBride}, {Fakhouri}, \&
  {Ma}}]{2009MNRAS.398.1858M}
{McBride}, J., {Fakhouri}, O., \& {Ma}, C.-P. 2009, \mnras, 398, 1858,
  \dodoi{10.1111/j.1365-2966.2009.15329.x}

\bibitem[{{Mihos} {et~al.}(2005){Mihos}, {Harding}, {Feldmeier}, \&
  {Morrison}}]{2005ApJ...631L..41M}
{Mihos}, J.~C., {Harding}, P., {Feldmeier}, J., \& {Morrison}, H. 2005, \apjl,
  631, L41, \dodoi{10.1086/497030}

\bibitem[{{Mink} \& {Kurtz}(1998)}]{1998ASPC..145...93M}
{Mink}, D.~J., \& {Kurtz}, M.~J. 1998, Astronomical Society of the Pacific
  Conference Series, Vol. 145, {RVSAO 2.0 - A Radial Velocity Package for
  IRAF}, ed. R.~{Albrecht}, R.~N. {Hook}, \& H.~A. {Bushouse}, 93

\bibitem[{{Mitchell} {et~al.}(2005){Mitchell}, {Keeton}, {Frieman}, \&
  {Sheth}}]{2005ApJ...622...81M}
{Mitchell}, J.~L., {Keeton}, C.~R., {Frieman}, J.~A., \& {Sheth}, R.~K. 2005,
  \apj, 622, 81, \dodoi{10.1086/427910}

\bibitem[{{Monna} {et~al.}(2015){Monna}, {Seitz}, {Zitrin}, {Geller}, {Grillo},
  {Mercurio}, {Greisel}, {Halkola}, {Suyu}, {Postman}, {Rosati}, {Balestra},
  {Biviano}, {Coe}, {Fabricant}, {Hwang}, \& {Koekemoer}}]{2015MNRAS.447.1224M}
{Monna}, A., {Seitz}, S., {Zitrin}, A., {et~al.} 2015, \mnras, 447, 1224,
  \dodoi{10.1093/mnras/stu2534}

\bibitem[{{Monna} {et~al.}(2017){Monna}, {Seitz}, {Geller}, {Zitrin},
  {Mercurio}, {Suyu}, {Postman}, {Fabricant}, {Hwang}, \&
  {Koekemoer}}]{2017MNRAS.465.4589M}
{Monna}, A., {Seitz}, S., {Geller}, M.~J., {et~al.} 2017, \mnras, 465, 4589,
  \dodoi{10.1093/mnras/stw3048}

\bibitem[{{Montero-Dorta} {et~al.}(2017){Montero-Dorta}, {Bolton}, \&
  {Shu}}]{2017MNRAS.468...47M}
{Montero-Dorta}, A.~D., {Bolton}, A.~S., \& {Shu}, Y. 2017, \mnras, 468, 47,
  \dodoi{10.1093/mnras/stx321}

\bibitem[{{Montero-Dorta} {et~al.}(2016){Montero-Dorta}, {Bolton},
  {Brownstein}, {Swanson}, {Dawson}, {Prada}, {Eisenstein}, {Maraston},
  {Thomas}, {Comparat}, {Chuang}, {McBride}, {Favole}, {Guo},
  {Rodr{\'\i}guez-Torres}, \& {Schneider}}]{2016MNRAS.461.1131M}
{Montero-Dorta}, A.~D., {Bolton}, A.~S., {Brownstein}, J.~R., {et~al.} 2016,
  \mnras, 461, 1131, \dodoi{10.1093/mnras/stw1352}

\bibitem[{{Moran} {et~al.}(2007){Moran}, {Ellis}, {Treu}, {Smith}, {Rich}, \&
  {Smail}}]{2007ApJ...671.1503M}
{Moran}, S.~M., {Ellis}, R.~S., {Treu}, T., {et~al.} 2007, \apj, 671, 1503,
  \dodoi{10.1086/522303}

\bibitem[{{Moster} {et~al.}(2010){Moster}, {Somerville}, {Maulbetsch}, {van den
  Bosch}, {Macci{\`o}}, {Naab}, \& {Oser}}]{2010ApJ...710..903M}
{Moster}, B.~P., {Somerville}, R.~S., {Maulbetsch}, C., {et~al.} 2010, \apj,
  710, 903, \dodoi{10.1088/0004-637X/710/2/903}

\bibitem[{{Munari} {et~al.}(2016){Munari}, {Grillo}, {De Lucia}, {Biviano},
  {Annunziatella}, {Borgani}, {Lombardi}, {Mercurio}, \&
  {Rosati}}]{2016ApJ...827L...5M}
{Munari}, E., {Grillo}, C., {De Lucia}, G., {et~al.} 2016, \apjl, 827, L5,
  \dodoi{10.3847/2041-8205/827/1/L5}

\bibitem[{{Murante} {et~al.}(2004){Murante}, {Arnaboldi}, {Gerhard}, {Borgani},
  {Cheng}, {Diaferio}, {Dolag}, {Moscardini}, {Tormen}, {Tornatore}, \&
  {Tozzi}}]{2004ApJ...607L..83M}
{Murante}, G., {Arnaboldi}, M., {Gerhard}, O., {et~al.} 2004, \apjl, 607, L83,
  \dodoi{10.1086/421348}

\bibitem[{{Naab} {et~al.}(2009){Naab}, {Johansson}, \&
  {Ostriker}}]{2009ApJ...699L.178N}
{Naab}, T., {Johansson}, P.~H., \& {Ostriker}, J.~P. 2009, \apjl, 699, L178,
  \dodoi{10.1088/0004-637X/699/2/L178}

\bibitem[{{Naab} {et~al.}(2006){Naab}, {Khochfar}, \&
  {Burkert}}]{2006ApJ...636L..81N}
{Naab}, T., {Khochfar}, S., \& {Burkert}, A. 2006, \apjl, 636, L81,
  \dodoi{10.1086/500205}

\bibitem[{{Napolitano} {et~al.}(2003){Napolitano}, {Pannella}, {Arnaboldi},
  {Gerhard}, {Aguerri}, {Freeman}, {Capaccioli}, {Ghigna}, {Governato},
  {Quinn}, \& {Stadel}}]{2003ApJ...594..172N}
{Napolitano}, N.~R., {Pannella}, M., {Arnaboldi}, M., {et~al.} 2003, \apj, 594,
  172, \dodoi{10.1086/376860}

\bibitem[{{Nipoti}(2017)}]{2017MNRAS.467..661N}
{Nipoti}, C. 2017, \mnras, 467, 661, \dodoi{10.1093/mnras/stx112}

\bibitem[{{Oogi} {et~al.}(2016){Oogi}, {Habe}, \&
  {Ishiyama}}]{2016MNRAS.456..300O}
{Oogi}, T., {Habe}, A., \& {Ishiyama}, T. 2016, \mnras, 456, 300,
  \dodoi{10.1093/mnras/stv2581}

\bibitem[{{Oser} {et~al.}(2010){Oser}, {Ostriker}, {Naab}, {Johansson}, \&
  {Burkert}}]{2010ApJ...725.2312O}
{Oser}, L., {Ostriker}, J.~P., {Naab}, T., {Johansson}, P.~H., \& {Burkert}, A.
  2010, \apj, 725, 2312, \dodoi{10.1088/0004-637X/725/2/2312}

\bibitem[{{Paccagnella} {et~al.}(2016){Paccagnella}, {Vulcani}, {Poggianti},
  {Moretti}, {Fritz}, {Gullieuszik}, {Couch}, {Bettoni}, {Cava}, {D'Onofrio},
  \& {Fasano}}]{2016ApJ...816L..25P}
{Paccagnella}, A., {Vulcani}, B., {Poggianti}, B.~M., {et~al.} 2016, \apjl,
  816, L25, \dodoi{10.3847/2041-8205/816/2/L25}

\bibitem[{{Papovich} {et~al.}(2018){Papovich}, {Kawinwanichakij}, {Quadri},
  {Glazebrook}, {Labb{\'e}}, {Tran}, {Forrest}, {Kacprzak}, {Spitler},
  {Straatman}, \& {Tomczak}}]{2018ApJ...854...30P}
{Papovich}, C., {Kawinwanichakij}, L., {Quadri}, R.~F., {et~al.} 2018, \apj,
  854, 30, \dodoi{10.3847/1538-4357/aaa766}

\bibitem[{{Park} {et~al.}(2007){Park}, {Choi}, {Vogeley}, {Gott}, {Blanton}, \&
  {SDSS Collaboration}}]{2007ApJ...658..898P}
{Park}, C., {Choi}, Y.-Y., {Vogeley}, M.~S., {et~al.} 2007, \apj, 658, 898,
  \dodoi{10.1086/511059}

\bibitem[{{Park} \& {Hwang}(2009)}]{2009ApJ...699.1595P}
{Park}, C., \& {Hwang}, H.~S. 2009, \apj, 699, 1595,
  \dodoi{10.1088/0004-637X/699/2/1595}

\bibitem[{{Peng} {et~al.}(2010){Peng}, {Lilly}, {Kova{\v{c}}}, {Bolzonella},
  {Pozzetti}, {Renzini}, {Zamorani}, {Ilbert}, {Knobel}, {Iovino}, {Maier},
  {Cucciati}, {Tasca}, {Carollo}, {Silverman}, {Kampczyk}, {de Ravel},
  {Sanders}, {Scoville}, {Contini}, {Mainieri}, {Scodeggio}, {Kneib}, {Le
  F{\`e}vre}, {Bardelli}, {Bongiorno}, {Caputi}, {Coppa}, {de la Torre},
  {Franzetti}, {Garilli}, {Lamareille}, {Le Borgne}, {Le Brun}, {Mignoli},
  {Perez Montero}, {Pello}, {Ricciardelli}, {Tanaka}, {Tresse}, {Vergani},
  {Welikala}, {Zucca}, {Oesch}, {Abbas}, {Barnes}, {Bordoloi}, {Bottini},
  {Cappi}, {Cassata}, {Cimatti}, {Fumana}, {Hasinger}, {Koekemoer},
  {Leauthaud}, {Maccagni}, {Marinoni}, {McCracken}, {Memeo}, {Meneux}, {Nair},
  {Porciani}, {Presotto}, \& {Scaramella}}]{2010ApJ...721..193P}
{Peng}, Y.-j., {Lilly}, S.~J., {Kova{\v{c}}}, K., {et~al.} 2010, \apj, 721,
  193, \dodoi{10.1088/0004-637X/721/1/193}

\bibitem[{{Pizzardo} {et~al.}(2020){Pizzardo}, {Di Gioia}, {Diaferio}, {De
  Boni}, {Serra}, {Geller}, {Sohn}, {Rines}, \& {Baldi}}]{2020arXiv200511562P}
{Pizzardo}, M., {Di Gioia}, S., {Diaferio}, A., {et~al.} 2020, arXiv e-prints,
  arXiv:2005.11562.
\newblock \doarXiv{2005.11562}

\bibitem[{{Ragone-Figueroa} {et~al.}(2018){Ragone-Figueroa}, {Granato},
  {Ferraro}, {Murante}, {Biffi}, {Borgani}, {Planelles}, \&
  {Rasia}}]{2018MNRAS.479.1125R}
{Ragone-Figueroa}, C., {Granato}, G.~L., {Ferraro}, M.~E., {et~al.} 2018,
  \mnras, 479, 1125, \dodoi{10.1093/mnras/sty1639}

\bibitem[{{Ragone-Figueroa} {et~al.}(2013){Ragone-Figueroa}, {Granato},
  {Murante}, {Borgani}, \& {Cui}}]{2013MNRAS.436.1750R}
{Ragone-Figueroa}, C., {Granato}, G.~L., {Murante}, G., {Borgani}, S., \&
  {Cui}, W. 2013, \mnras, 436, 1750, \dodoi{10.1093/mnras/stt1693}

\bibitem[{{Rines} \& {Diaferio}(2006)}]{2006AJ....132.1275R}
{Rines}, K., \& {Diaferio}, A. 2006, \aj, 132, 1275, \dodoi{10.1086/506017}

\bibitem[{{Rines} {et~al.}(2013){Rines}, {Geller}, {Diaferio}, \&
  {Kurtz}}]{2013ApJ...767...15R}
{Rines}, K., {Geller}, M.~J., {Diaferio}, A., \& {Kurtz}, M.~J. 2013, \apj,
  767, 15, \dodoi{10.1088/0004-637X/767/1/15}

\bibitem[{{Rines} {et~al.}(2016){Rines}, {Geller}, {Diaferio}, \&
  {Hwang}}]{2016ApJ...819...63R}
{Rines}, K.~J., {Geller}, M.~J., {Diaferio}, A., \& {Hwang}, H.~S. 2016, \apj,
  819, 63, \dodoi{10.3847/0004-637X/819/1/63}

\bibitem[{{Rines} {et~al.}(2018){Rines}, {Geller}, {Diaferio}, {Hwang}, \&
  {Sohn}}]{2018ApJ...862..172R}
{Rines}, K.~J., {Geller}, M.~J., {Diaferio}, A., {Hwang}, H.~S., \& {Sohn}, J.
  2018, \apj, 862, 172, \dodoi{10.3847/1538-4357/aacd49}

\bibitem[{{Rodriguez-Gomez} {et~al.}(2016){Rodriguez-Gomez}, {Pillepich},
  {Sales}, {Genel}, {Vogelsberger}, {Zhu}, {Wellons}, {Nelson}, {Torrey},
  {Springel}, {Ma}, \& {Hernquist}}]{2016MNRAS.458.2371R}
{Rodriguez-Gomez}, V., {Pillepich}, A., {Sales}, L.~V., {et~al.} 2016, \mnras,
  458, 2371, \dodoi{10.1093/mnras/stw456}

\bibitem[{{S{\'a}nchez-Bl{\'a}zquez} {et~al.}(2006){S{\'a}nchez-Bl{\'a}zquez},
  {Gorgas}, {Cardiel}, \& {Gonz{\'a}lez}}]{2006A&A...457..809S}
{S{\'a}nchez-Bl{\'a}zquez}, P., {Gorgas}, J., {Cardiel}, N., \& {Gonz{\'a}lez},
  J.~J. 2006, \aap, 457, 809, \dodoi{10.1051/0004-6361:20064845}

\bibitem[{{Serra} \& {Diaferio}(2013)}]{2013ApJ...768..116S}
{Serra}, A.~L., \& {Diaferio}, A. 2013, \apj, 768, 116,
  \dodoi{10.1088/0004-637X/768/2/116}

\bibitem[{{Serra} {et~al.}(2011){Serra}, {Diaferio}, {Murante}, \&
  {Borgani}}]{2011MNRAS.412..800S}
{Serra}, A.~L., {Diaferio}, A., {Murante}, G., \& {Borgani}, S. 2011, \mnras,
  412, 800, \dodoi{10.1111/j.1365-2966.2010.17946.x}

\bibitem[{{Sheth} {et~al.}(2003){Sheth}, {Bernardi}, {Schechter}, {Burles},
  {Eisenstein}, {Finkbeiner}, {Frieman}, {Lupton}, {Schlegel}, {Subbarao},
  {Shimasaku}, {Bahcall}, {Brinkmann}, \& {Ivezi{\'c}}}]{2003ApJ...594..225S}
{Sheth}, R.~K., {Bernardi}, M., {Schechter}, P.~L., {et~al.} 2003, \apj, 594,
  225, \dodoi{10.1086/376794}

\bibitem[{{Sohn} {et~al.}(2020){Sohn}, {Geller}, {Diaferio}, \&
  {Rines}}]{2020ApJ...891..129S}
{Sohn}, J., {Geller}, M.~J., {Diaferio}, A., \& {Rines}, K.~J. 2020, \apj, 891,
  129, \dodoi{10.3847/1538-4357/ab6e6a}

\bibitem[{{Sohn} {et~al.}(2019{\natexlab{a}}){Sohn}, {Geller}, {Walker},
  {Dell'Antonio}, {Diaferio}, \& {Rines}}]{2019ApJ...871..129S}
{Sohn}, J., {Geller}, M.~J., {Walker}, S.~A., {et~al.} 2019{\natexlab{a}},
  \apj, 871, 129, \dodoi{10.3847/1538-4357/aaf1cc}

\bibitem[{{Sohn} {et~al.}(2019{\natexlab{b}}){Sohn}, {Geller}, {Zahid}, \&
  {Fabricant}}]{2019ApJ...872..192S}
{Sohn}, J., {Geller}, M.~J., {Zahid}, H.~J., \& {Fabricant}, D.~G.
  2019{\natexlab{b}}, \apj, 872, 192, \dodoi{10.3847/1538-4357/ab0213}

\bibitem[{{Sohn} {et~al.}(2017{\natexlab{a}}){Sohn}, {Geller}, {Zahid},
  {Fabricant}, {Diaferio}, \& {Rines}}]{2017ApJS..229...20S}
{Sohn}, J., {Geller}, M.~J., {Zahid}, H.~J., {et~al.} 2017{\natexlab{a}},
  \apjs, 229, 20, \dodoi{10.3847/1538-4365/aa653e}

\bibitem[{{Sohn} {et~al.}(2017{\natexlab{b}}){Sohn}, {Zahid}, \&
  {Geller}}]{2017ApJ...845...73S}
{Sohn}, J., {Zahid}, H.~J., \& {Geller}, M.~J. 2017{\natexlab{b}}, \apj, 845,
  73, \dodoi{10.3847/1538-4357/aa7de3}

\bibitem[{{Springel} {et~al.}(2018){Springel}, {Pakmor}, {Pillepich},
  {Weinberger}, {Nelson}, {Hernquist}, {Vogelsberger}, {Genel}, {Torrey},
  {Marinacci}, \& {Naiman}}]{2018MNRAS.475..676S}
{Springel}, V., {Pakmor}, R., {Pillepich}, A., {et~al.} 2018, \mnras, 475, 676,
  \dodoi{10.1093/mnras/stx3304}

\bibitem[{{Thomas} {et~al.}(2013){Thomas}, {Steele}, {Maraston}, {Johansson},
  {Beifiori}, {Pforr}, {Str{\"o}mb{\"a}ck}, {Tremonti}, {Wake}, {Bizyaev},
  {Bolton}, {Brewington}, {Brownstein}, {Comparat}, {Kneib}, {Malanushenko},
  {Malanushenko}, {Oravetz}, {Pan}, {Parejko}, {Schneider}, {Shelden},
  {Simmons}, {Snedden}, {Tanaka}, {Weaver}, \& {Yan}}]{2013MNRAS.431.1383T}
{Thomas}, D., {Steele}, O., {Maraston}, C., {et~al.} 2013, \mnras, 431, 1383,
  \dodoi{10.1093/mnras/stt261}

\bibitem[{{Tonini} {et~al.}(2012){Tonini}, {Bernyk}, {Croton}, {Maraston}, \&
  {Thomas}}]{2012ApJ...759...43T}
{Tonini}, C., {Bernyk}, M., {Croton}, D., {Maraston}, C., \& {Thomas}, D. 2012,
  \apj, 759, 43, \dodoi{10.1088/0004-637X/759/1/43}

\bibitem[{{Tormen} {et~al.}(1998){Tormen}, {Diaferio}, \&
  {Syer}}]{1998MNRAS.299..728T}
{Tormen}, G., {Diaferio}, A., \& {Syer}, D. 1998, \mnras, 299, 728,
  \dodoi{10.1046/j.1365-8711.1998.01775.x}

\bibitem[{{Utsumi} {et~al.}(2020){Utsumi}, {Geller}, {Zahid}, {Sohn},
  {Dell'Antonio}, {Kawanomoto}, {Komiyama}, {Koshida}, \&
  {Miyazaki}}]{2020arXiv200507122U}
{Utsumi}, Y., {Geller}, M.~J., {Zahid}, H.~J., {et~al.} 2020, arXiv e-prints,
  arXiv:2005.07122.
\newblock \doarXiv{2005.07122}

\bibitem[{{Vale} \& {Ostriker}(2004)}]{2004MNRAS.353..189V}
{Vale}, A., \& {Ostriker}, J.~P. 2004, \mnras, 353, 189,
  \dodoi{10.1111/j.1365-2966.2004.08059.x}

\bibitem[{{van den Bosch}(2017)}]{2017MNRAS.468..885V}
{van den Bosch}, F.~C. 2017, \mnras, 468, 885, \dodoi{10.1093/mnras/stx520}

\bibitem[{{van den Bosch} {et~al.}(2014){van den Bosch}, {Jiang}, {Hearin},
  {Campbell}, {Watson}, \& {Padmanabhan}}]{2014MNRAS.445.1713V}
{van den Bosch}, F.~C., {Jiang}, F., {Hearin}, A., {et~al.} 2014, \mnras, 445,
  1713, \dodoi{10.1093/mnras/stu1872}

\bibitem[{{van den Bosch} {et~al.}(2018){van den Bosch}, {Ogiya}, {Hahn}, \&
  {Burkert}}]{2018MNRAS.474.3043V}
{van den Bosch}, F.~C., {Ogiya}, G., {Hahn}, O., \& {Burkert}, A. 2018, \mnras,
  474, 3043, \dodoi{10.1093/mnras/stx2956}

\bibitem[{{van Uitert} {et~al.}(2013){van Uitert}, {Hoekstra}, {Franx},
  {Gilbank}, {Gladders}, \& {Yee}}]{2013A&A...549A...7V}
{van Uitert}, E., {Hoekstra}, H., {Franx}, M., {et~al.} 2013, \aap, 549, A7,
  \dodoi{10.1051/0004-6361/201220439}

\bibitem[{{Vulcani} {et~al.}(2013){Vulcani}, {Poggianti}, {Oemler}, {Dressler},
  {Arag{\'o}n-Salamanca}, {De Lucia}, {Moretti}, {Gladders}, {Abramson}, \&
  {Halliday}}]{2013A&A...550A..58V}
{Vulcani}, B., {Poggianti}, B.~M., {Oemler}, A., {et~al.} 2013, \aap, 550, A58,
  \dodoi{10.1051/0004-6361/201118388}

\bibitem[{{Wake} {et~al.}(2012){Wake}, {van Dokkum}, \&
  {Franx}}]{2012ApJ...751L..44W}
{Wake}, D.~A., {van Dokkum}, P.~G., \& {Franx}, M. 2012, \apjl, 751, L44,
  \dodoi{10.1088/2041-8205/751/2/L44}

\bibitem[{{Wechsler} \& {Tinker}(2018)}]{2018ARA&A..56..435W}
{Wechsler}, R.~H., \& {Tinker}, J.~L. 2018, \araa, 56, 435,
  \dodoi{10.1146/annurev-astro-081817-051756}

\bibitem[{{Yang} {et~al.}(2008){Yang}, {Mo}, \& {van den
  Bosch}}]{2008ApJ...676..248Y}
{Yang}, X., {Mo}, H.~J., \& {van den Bosch}, F.~C. 2008, \apj, 676, 248,
  \dodoi{10.1086/528954}

\bibitem[{{Zabludoff} \& {Mulchaey}(1998)}]{1998ApJ...496...39Z}
{Zabludoff}, A.~I., \& {Mulchaey}, J.~S. 1998, \apj, 496, 39,
  \dodoi{10.1086/305355}

\bibitem[{{Zahid} \& {Geller}(2017)}]{2017ApJ...841...32Z}
{Zahid}, H.~J., \& {Geller}, M.~J. 2017, \apj, 841, 32,
  \dodoi{10.3847/1538-4357/aa7056}

\bibitem[{{Zahid} {et~al.}(2016){Zahid}, {Geller}, {Fabricant}, \&
  {Hwang}}]{2016ApJ...832..203Z}
{Zahid}, H.~J., {Geller}, M.~J., {Fabricant}, D.~G., \& {Hwang}, H.~S. 2016,
  \apj, 832, 203, \dodoi{10.3847/0004-637X/832/2/203}

\bibitem[{{Zahid} {et~al.}(2018){Zahid}, {Sohn}, \&
  {Geller}}]{2018ApJ...859...96Z}
{Zahid}, H.~J., {Sohn}, J., \& {Geller}, M.~J. 2018, \apj, 859, 96,
  \dodoi{10.3847/1538-4357/aabe31}

\bibitem[{{Zhao} {et~al.}(2009){Zhao}, {Jing}, {Mo}, \&
  {B{\"o}rner}}]{2009ApJ...707..354Z}
{Zhao}, D.~H., {Jing}, Y.~P., {Mo}, H.~J., \& {B{\"o}rner}, G. 2009, \apj, 707,
  354, \dodoi{10.1088/0004-637X/707/1/354}

\end{thebibliography}
\bibliographystyle{aasjournal}

\end{document}